\documentclass{aa}
\usepackage{graphicx}
\usepackage{txfonts}
\usepackage{natbib}
\bibpunct{(}{)}{;}{a}{}{,}
\newcommand\lineone{\ion{Fe}{i}~$\lambda$6301.5~\AA}
\newcommand\linetwo{\ion{Fe}{i}~$\lambda$6302.5~\AA}
\newcommand\linethree{O$_{2}$~$\lambda$6302.8~\AA}

\begin{document}

\title{
	Inter-Network magnetic fields 
        \\ observed with sub-arcsec resolution
}	

  \author{I. Dom\'\i nguez Cerde\~na\inst{1}
          \and
          J. S\'anchez Almeida\inst{2}
          \and
          F. Kneer\inst{1}
	  }
 \offprints{I. Dom\'\i nguez Cerde\~na}

   \mail{ita@uni-sw.gwdg.de}

   \institute{Universit\"ats-Sternwarte,
              Geismarlandstra\ss e 11, D-37083 G\"ottingen, Germany
   \and
             Instituto de Astrof\'\i sica de Canarias, 
              E-38200 La Laguna, Spain}

   \date{Received date / Accepted date}

\abstract{We analyze a time sequence 
of Inter-Network (IN) magnetograms
observed at the solar disk center.
Speckle reconstruction techniques provide 
a good
spatial resolution (0\farcs5 cutoff frequency) yet maintaining a fair sensitivity
(some 20\,G). Patches with signal above noise 
cover 60\,\% of the observed area, most of which 
corresponds to intergranular lanes.
The large surface covered by signal renders a mean
unsigned magnetic flux density between 17\,G 
and 21\,G (1\,G$\equiv$~1~Mx cm$^{-2}$). The
difference depends on the spectral line
used to generate the magnetograms (\linetwo\ or \lineone ). Such
systematic difference can be understood if the magnetic
structures 
producing the polarization have intrinsic field strengths
exceeding 1 kG, and consequently, 
occupying only a very small
fraction of the surface (some 2\%).
We observe both, magnetic signals changing in
time scales smaller than 1 min,
and a persistent pattern lasting longer than the
duration of the sequence (17 min). The pattern resembles
a network with a spatial scale between 5  and 10 arcsec, which
we identify as the mesogranulation.
The strong dependence of the polarization signals 
on spatial resolution and sensitivity suggests that 
much quiet Sun magnetic flux still remains undetected. 
}


%
	\authorrunning{Dom\'\i nguez Cerde\~na et al.}
	\titlerunning{Inter-Network magnetic fields with sub-arcsec resolution}

\maketitle	

\section{Introduction}

Most of the  solar surface appears as {\em non-magnetic}
in traditional magnetic field 
determinations 
\citep[e.g., in the full-disk Kitt Peak 
magnetograms;][]{jon92}.
This so-called {\em quiet Sun} does not produce enough 
polarization to show up in the  measurements. 
However, such lack of detection does not imply the 
non-existence or irrelevance of the quiet Sun
magnetism.
On the contrary, the limited  sensitivity of the standard
magnetograms and the large area covered by 
the quiet Sun point out that traditional 
measurements may easily overlook a large 
fraction of the solar magnetic flux.
With various flavors and shades, this  argument 
has been put forward many times
during the past fifty years 
	\citep[e.g.][]
	{unn59,
	ste82, 
	zir87,
	yi93,
	san98c, 
	san02b}.
If the conjecture were correct and the quiet Sun
carries a sizeable fraction of the  existing
magnetic flux, then
weak polarization signals should appear upon improvement
of the sensitivity of the magnetograms. 
Such weak signals are actually
observed thanks to the last generation of solar 
spectro-polarimeters.
When the noise
is in the few G level and the angular resolution about 1\arcsec ,
then most of the solar surface becomes magnetic
\citep[e.g.,][]{gro96,lin99,lit02}. 
In addition, 
Hanle depolarization
measurements of chromospheric lines 
also indicate
the presence of an ubiquitous 
magnetic field 
\citep[e.g.,][]{fau95, bia99,shc03}.

Little is known about the physical properties of these fields
(magnetic flux, distribution of field strengths, structure,
degree of concentration and tangling, relationship with the fields
in the chromosphere and corona, etc.). The observational studies
are still exploratory, aiming at setting up the scene.
With this general purpose, we have observed the  
magnetic fields of an Inter-Network (IN) region
with good angular resolution, yet maintaining a fair sensitivity. The study refers
to low flux features in the interior of  
a network cell, excluding network patches for which 
an extensive literature exists \citep[e.g.,][]{sol93,ste94}. 
IN magnetic fields were
already discovered in the seventies \citep{liv75,smi75}.
During the last decade, with the new observational and diagnostic
capabilities, these IN fields have received increasing attention. 
So far only a fraction of the IN field has been detected. This conclusion
follows from the presence of unresolved mixed polarities in  
1\arcsec\ angular resolution  observations
\citep[e.g.,][]{san96,san00,lit02}. The mixing
of polarities reduces the polarization,
making the IN magnetic structures difficult to identify.
It is therefore to be expected that more locations with magnetic
fields will be detected when increasing the spatial resolution. 
Some of the recent observational studies find IN fields having
magnetic field strengths substantially lower than 1~kG
	\citep{kel94,
	lin95,
	lin99,
	bia98,
	bia99,
	kho02}.
On the contrary,
different observing and interpretation techniques indicate
the existence of strong kG magnetic 
fields \citep{gro96,
	sig99,
	san00,
	soc02}.
The inconsistency could be cured 
if the IN regions present a continuous distribution of field strengths. 
Depending on the specificities of the diagnostic technique,
one is mostly
sensitive to a particular part of such distribution \citep[see][]{cat99a,san00,soc03}.
Lifetimes of IN features have been measured to be 
between 0.2 to 7.5 hours \citep{zha98}. 
Improving 
of the angular resolution and cadence reduces the lifetimes, 
which come down to a few minutes \citep{lin99}. 
Such combination of short lifetimes with the large flux content
makes IN regions a very efficient system to process magnetic fields,
orders of magnitude more effective than both
active regions 
\citep[uplifting $~7\,\times\,10^{21}$~Mx~day$^{-1}$
during the maximum of the cycle,][]{har93}
and ephemeral regions 
\citep[$\sim\,5\,\times\,10^{23}$~Mx~day$^{-1}$,][]{hag01}.

Here we measure several basic physical properties
of the IN fields. The novel side of the study lies in
the  good spatial resolution of the time series
of magnetograms 
on which the measurements are based (0\farcs5).
The observations are presented in Sect. \ref{observations}.
Then Sect. \ref{data} describes how the high spatial
resolution is obtained by speckle reconstruction,
and how the magnetograms are produced from the
reconstructed spectro-polarimetric maps.
We
analyze in detail the best snapshot of the
sequence to characterize
the  properties of the IN fields (Sect.~\ref{results}). 
The complete series allows us 
to study the time evolution of the magnetic signals
(Sect. \ref{sectemp}). 
In order to assess the reliability of our conclusions,
we analyze in detail the consistency of the 
results, both internally and when compared with previous
measurements (Sect. \ref{consistency}).
Some of the results presented here
were already advanced in a letter by
\citet{dom03}.


\section{Observations\label{observations}}

\begin{figure}
\resizebox{\hsize}{!}{\includegraphics{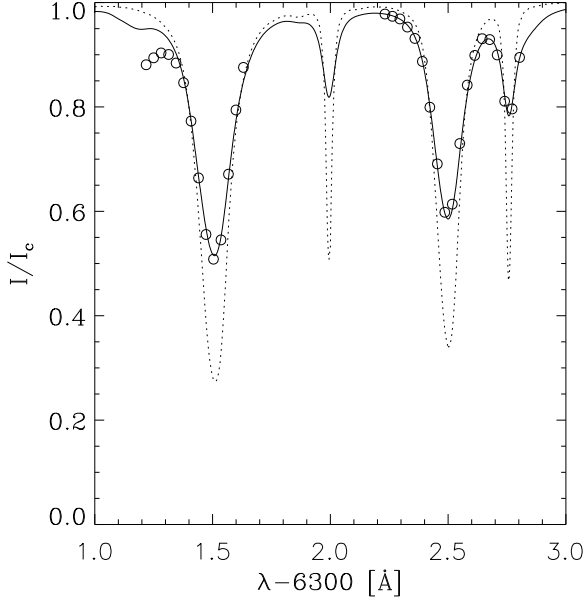}}   
\caption{Comparison between a Stokes $I$ profile from our flat field data
(open circles) and a reference solar spectrum
before (dotted line) and after (solid line)  
convolution with the FPI band-pass function (FWHM~44~m\AA).
The wavelength $\lambda$ is referred to 6300\,\AA , and 
the Stokes $I$ profiles are normalized to the continuum intensity $I_c$.
}
\label{flat} 
\end{figure}

The observations were obtained with the G\"ottingen
Fabry-Perot Interferometer (FPI),
which  is  a post-focus instrument 
of the German Vacuum Tower Telescope
of the Observatorio del Teide (Tenerife, Spain). 
The target was a 
very quiet Sun region near disk center,
which was selected using G-band video images to avoid contamination
from network magnetic concentrations\footnote{Further
evidence showing that the data belong to an
IN region will be presented below.}.
The optical system derives from the early work by \citet{ben92},
\citet{ben93}, and \citet{ben95}. Its present setup is described
in \citet{kos01}.
%
It includes two CCD cameras, operating simultaneously,
with an exposure time of 30~ms. 
Both CCDs have 384$\times$286 pixels, each pixel being 
0\farcs 1~$\times$~0\farcs 1 on the Sun, which corresponds to
half  the diffraction limit of the telescope at the working 
wavelength.
The broad-band camera (CCD1)
images the field-of-view (FOV) through
an interference filter 
(band-pass $\sim\,$100~\AA, 
centered at $\sim\,$6300~\AA).
It provides the so-called speckle image,
needed for the speckle reconstruction (Sect. \ref{secsp}).
The second camera (CCD2) gathers narrow-band images whose
wavelength is selected by the FPI. This second
beam includes 
a Stokes $V$  analyzer, which separates 
the left and right circularly polarized
components of the light
into different areas of the CCD2
\citep[for details, see Fig. 1
in][]{kos01, kos01b}.
The FPI allows to scan in wavelength, taking several images
per wavelength position. We select a spectral region around $\lambda$6302~\AA ,
which contains the iron lines \lineone~(Land\'e factor
$g_L=1.67$), \linetwo~($g_L=2.5$), as well as 
the telluric line \linethree .
The FPI wavelength step was set to
31.8~m\AA , which suffices to sample the telluric line (5 wavelengths,
with 2 images per position),
and the two \ion{Fe}{i} lines (14 positions each, with 5 images
per position).
The dots in Fig.~\ref{flat} indicate the exact  wavelengths. 
A full scan with the properties described above
renders 150 images, which is the maximum  number
allowed by the acquisition system.
Each scan lasts 35~s, plus 15~s needed for storage onto hard disk. 

The FPI was adjusted to yield a band-pass 
of 44~m\AA~FWHM (Full Width Half Maximum). 
This finite wavelength resolution modifies the spectrum.
Figure \ref{flat} shows the mean Stokes $I$  profile 
in one of our flatfields, together with the spectrum of the 
disk center as published in a standard solar atlas
\citep[Brault \& Neckel 1987, quoted in][]{nec99}. Note the good agreement between 
our spectrum and
the atlas smeared with the theoretical band-pass of the FPI.

The data used in the work belong 
to a short time series of 20 scans taken
in April 29, 2002. 
The cadence was some 50~s, so that the
series spans 17~min. The 
seeing conditions were particularly good, with a Fried
parameter deduced from the speckle reconstruction
between $13$ and $14$~cm. 

\section{Data analysis\label{data}}

\subsection{Speckle reconstruction of the narrow-band images\label{secsp}}
\begin{figure}
\resizebox{\hsize}{!}{\includegraphics{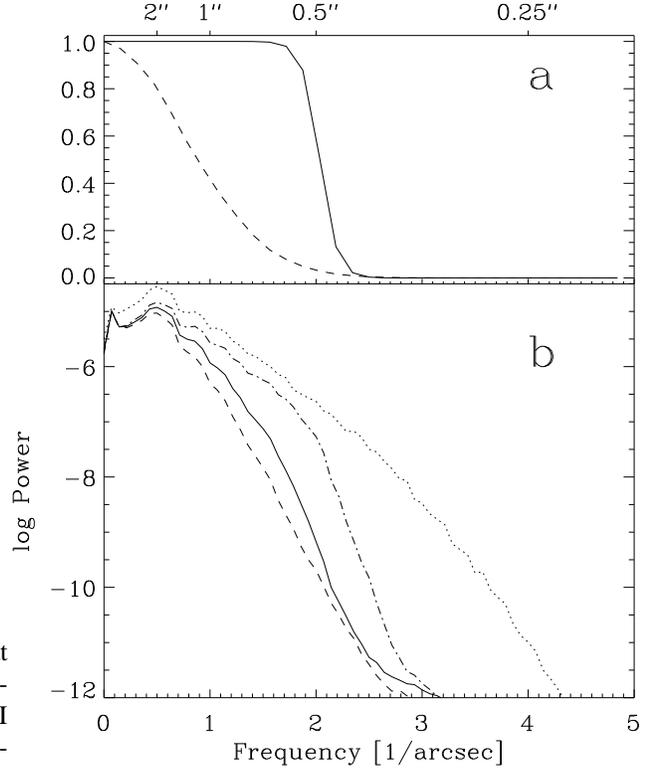}}   
\caption{
Filters and power spectra involved in the process
of image restoration. {\bf a)} Solid line: noise filter applied
to the series of individual frames that give rise to each
restored image. Dashed line: optical
transfer function  for a long exposure image with 
0\farcs5 FWHM seeing.
{\bf b)} Dotted line: power spectrum of the broad-band image.
Dash-dotted line: power spectrum of the narrow-band continuum image 
processed with the noise filter.
Solid line: the narrow-band continuum image 
after smoothing. Dashed line:  
power spectrum of the narrow-band continuum image
considering long exposure seeing.
Frequencies are given in 1/arcsec. The corresponding
periods are also shown on the upper bound  of the plot.
} 
\label{pow}
\end{figure}

The reconstruction of the FPI narrow-band images is a two-step process,
which begins by reconstructing the broad-band images.
Using the 150 broad-band images taken in CCD1, one is able to obtain a
reconstructed image with a spatial resolution close to
0\farcs25 . The G\"ottingen code for speckle reconstruction is
used as described in \citet{deb92b}  and \citet{deb96}.
We use the spectral ratio method \citep{von84} to derive the
seeing conditions and the amplitude correction factors.
The speckle masking method \citep{wei77,wei83}
then provides accurate phases at high spatial frequencies.
The narrow-band image reconstruction was carried 
out using the code by \citet{jan03}, which we employ after
minor modifications.
This code implements the method of \citet{kel92b}.
The instantaneous Optical 
Transfer Function is obtained 
from the instantaneous broad-band images assuming that the 
broad-band speckle reconstructed image is the true scenery.
Its Fourier transform is denoted by $\hat O_b$, where
the hat indicates that we deal with an estimate.
Under this assumption, one can write down an equation that
yields the Fourier transform of the reconstructed
narrow-band image $\hat O_{n}$ as a function of known quantities,
namely, $\hat O_b$, 
and the Fourier 
transforms of the individual 
broad-band ${I}_{b,j}$ and narrow-band ${I}_{n,j}$ images,
\begin{equation}
\hat O_{n}=H
\frac{\sum_{j}{I}_{n,j}{I}_{b,j}^{\ast}}{\sum_{j}|{I}_{b,j}|^{2}}\hat O_{b}.
\label{eqspec}
\end{equation}
The sums comprise all the images taken for a given wavelength,
$H$ 
represents a noise filter, and the superscript $\ast$
denotes complex conjugate. 
All the symbols in Eq. (\ref{eqspec}) are
functions of the spatial frequency (not explicitly included).
Equation (\ref{eqspec}) provides the base for the
reconstruction technique.
It derives, apart from the noise filter $H$, from a least-squares fit
that minimizes the merit function
\begin{equation}
\vert E\vert^2=
\sum_{j}(\hat O_n S_j-I_{n,j})\cdot(\hat O_n^{\ast}S_j^{\ast}-I_{n,j}^{\ast}),
\label{leastsqu}
\end{equation}
with respect to $\hat O_{n}$.
The instantaneous Optical Transfer Function $S_j$ has to be 
replaced by
\begin{equation}
S_j=\frac{I_{bj}}{\hat O_b};
\label{iotf}
\end{equation}
see \citet{kel92b}, \citet{kri99}, or \citet{kos01}.

The noise filter $H$ is usually an optimum filter 
calculated according to the signal 
\citep[e.g.,][]{kri99}. However, the signal depends on the
wavelength within the spectral line, 
which would produce a noise filter varying  along the wavelength scan
and, consequently, a spatial resolution different for different
wavelengths.
In this study we need to assure the same 
resolution for all wavelengths since the reconstructed images 
at different wavelengths are
combined to form individual spectra. For this reason  
we use a single noise filter which  
equals one until the spatial
frequency corresponding to 0\farcs5, and then it 
drops to zero from this point on (see Fig.~\ref{pow}).
The 0\farcs5\ cutoff was selected 
because
it is not far from the values 
provided by the optimum filter, and 
it still renders a good angular resolution.
To further suppress the noise, the
final narrow-band images were smoothed with a $5\times 5$ pixels boxcar.
Figure \ref{pow} shows azimuthal averages of various
power spectra that take part in the process of
speckle restoration. Note, in particular, how the broad-band image
has power up to a spatial scale equivalent to 0\farcs25, whereas
the corresponding power spectrum of the narrow-band image drops off 
beyond 0\farcs5.
\begin{figure*}[thb]
\centering
\includegraphics[width=17cm]{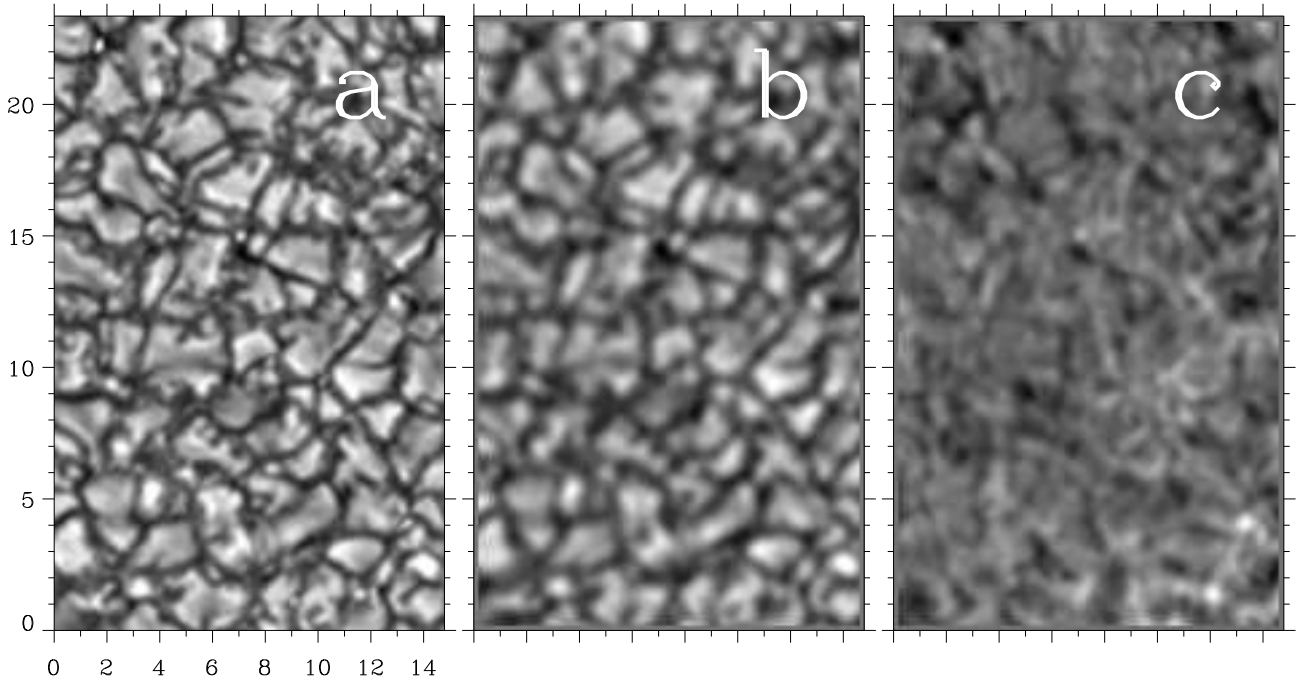}
\caption{Example of reconstruction. {\bf a)} Broad-band image (spatial resolution
$\sim$~0\farcs 25).
{\bf b)} narrow-band continuum 
image (spatial resolution $\sim$~0\farcs 5). {\bf c)} Intensity at the
core of \linetwo. The axes are in arcsec.
} 
\label{imgs} 
\end{figure*} 
Figure \ref{imgs} shows a reconstruction. The axes are
in arcsec so the FOV corresponds to 14\arcsec~$\times$~23\arcsec . 
Figure \ref{imgs}a contains the broad-band image. The contrast of this
image turns out to be 12.5\% (by definition, the contrast is the standard
deviation of the intensity in the image normalized to the mean intensity).
Figure~\ref{imgs}b shows the continuum intensity of the
narrow-band reconstruction. In this case the contrast drops to
some 7\%. Figure~\ref{imgs}c is the intensity at the core of
\linetwo. 
Incidentally, the latter image provides yet 
another evidence supporting that the observations
correspond to a very
quiet Sun region. 
Network concentrations show up as conspicuous
bright points at line core images \citep[e.g.,][]{she67}.

\subsection{Stokes $I$ and Stokes $V$ profiles\label{profiles}}
\begin{figure}
\resizebox{0.9\hsize}{!}{\includegraphics[trim=-5cm -2cm 0cm 0cm]{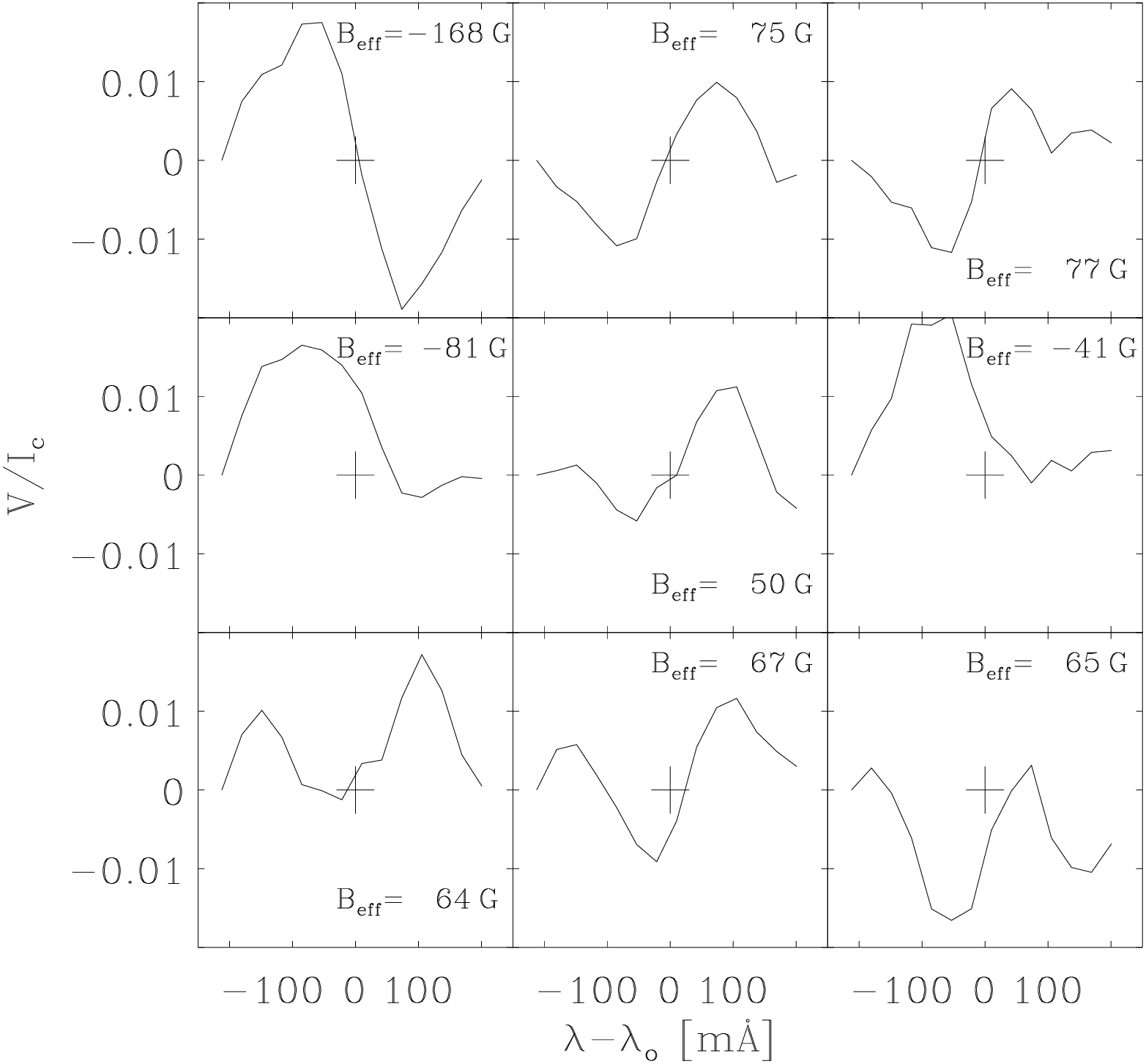}}
\caption{
Set of representative Stokes $V$ profiles of \linetwo. Despite the noise, deviations
	from the anti-symmetric shape are evident.
	(The large plus signs point out the origin of abscissae
	and ordinates.)
	We do not know which
	fraction of these asymmetries is real and which results
	from contamination with Stokes $I$. 
	Wavelengths are in m\AA\ off line core $\lambda_0$.
	The polarization signals have been normalized to the continuum
	intensity $I_c$. The flux density assigned to each profile
	is included for reference. 
}
\label{asymmetries}
\end{figure}

The restoration process is carried out independently for the
two images corresponding to the right and left
circularly polarized beams $S_1$ and $S_2$. 
Once the restoration is completed, we compute the
Stokes $I$ and Stokes $V$ profiles\footnote{The term 
{\em Stokes profiles} denotes the variation with wavelength
of the four Stokes parameters. In the present work we only deal with the
intensity $I$ and the circular polarization $V$.} 
by adding and subtracting
the two images at each wavelength,
\begin{eqnarray}
V&\propto S_1-S_2,\cr
I&\propto S_1+S_2.
	\label{profiles1}
\end{eqnarray}
This task demands a careful superposition of the beams, which
we carry out after sub-pixel interpolation.
An example of a Stokes $I$ profile is included in Fig. \ref{flat}.
Several Stokes $V$ profiles are shown in Fig. \ref{asymmetries}.

\subsection{Magnetograms and their calibration\label{magcal}}

	Magnetic flux densities are computed starting from the 
the well-known magnetograph equation, which
relates the circular polarization $V$ and
the longitudinal component
of the magnetic field $B$ \citep[e.g.,][]{unn56,lan92},
\begin{equation} 
V(\lambda) = C(\lambda)\ B, 
\label{eeqmag} 
\end{equation} 
with 
\begin{equation}
C(\lambda)=k g_{L} \lambda_0^2 \frac{dI(\lambda)}{d\lambda}.
\label{g_eff} 
\end{equation} 
The symbols $\lambda$,  $\lambda_0$ and $g_L$ stand for the wavelength,
the central wavelength of the line, and the
effective Land\'e factor, respectively. 
When Stokes $V$ and $I$ have the same units,
the constant $k$ is equal to 
$-~4.67~\cdot~10^{-13}$~\AA$^{-1}$~G$^{-1}$.
We apply an independent calibration
to each position of the
field of view.
First, the
derivative of Stokes $I$ required to evaluate the
calibration constant $C(\lambda)$ is computed numerically.
Then, we select the wavelengths of the two extrema of the 
derivative, plus the two wavelengths immediately adjacent to 
each one of them. The selection renders six independent 
wavelength positions with $C(\lambda)\not= 0$ where Eq. 
(\ref{eeqmag})
could be applied to estimate $B$. Each wavelength
gives a different value, so we chose a {\em best estimate} 
solving Eq. (\ref{eeqmag}) by
lest-squares. This procedure
defines our estimate of the
flux density $B_\mathrm{eff}$,
\begin{equation}
B_\mathrm{eff}=\frac{\sum_j V(\lambda_j) C(\lambda_j)}{\sum_j C^{2}(\lambda_j)}.
\label{llest}
\end{equation} 
The sum considers the six wavelengths
$\lambda_j$ whose selection is described above
(see also the square symbols in Fig. \ref{example}). Obviously, if 
the observed Stokes $I$ and $V$ profiles follow Eq.~(\ref{eeqmag}) 
then $B_\mathrm{eff}=B$. 
In practice this is not the case, and an extensive 
literature discusses various bias resulting from the
break down of the magnetograph equation \citep[refer to, e.g.,][]{jef91,kel94,gra02}. 
Here we need to
consider the main bias arising when the
magnetic structure is much smaller than the resolution 
element, mostly filled by unmagnetized  plasma.
In this case,
\begin{equation}
B_\mathrm{eff}\simeq \alpha B,
\label{fill_factor}
\end{equation}
$\alpha$ being the fraction of resolution
element occupied by the otherwise uniform magnetic
structure. According to Eq. (\ref{fill_factor}), 
$B_\mathrm{eff}$ turns out to estimate 
the magnetic flux across the resolution element
divided by the area of the resolution element (i.e., the
magnetic flux density).

	Two main sources of noise limit the precision of the 
flux density derived from Eq. (\ref{llest}).
The measure is affected by the random noise of the 
Stokes $V$ spectra.
We estimate its influence applying the law of propagation of errors 
\citep[e.g.,][]{mar71} to Eq. (\ref{llest}), i.e.,
\begin{displaymath}
\Delta B_\mathrm{eff}^2\simeq\sum_j\big[{\partial B_\mathrm{eff}}/{\partial V(\lambda_j)}\big]^2 
		\Delta V(\lambda_j)^2,
\end{displaymath}
\begin{equation}
~~~~~~~~\simeq \Delta V^2/\sum_j C^{2}(\lambda_j),
\label{cal0}
\end{equation}
where one assumes  that the  noise of the different wavelengths
$\Delta V(\lambda_j)$
is independent but has the same variance $\Delta V^2$.
We have applied the previous equation to all individual profiles.
The mean values of the errors are
\begin{eqnarray}
\Delta B_\mathrm{eff}\simeq& 23\ {\rm G\ for\ Fe}\ {\sc I}\ \lambda 6301.5\ {\rm \AA},\cr
\Delta B_\mathrm{eff}\simeq& 17\ {\rm G\ for\ Fe}\ {\sc I}\ \lambda 6302.5\ {\rm \AA}.
	\label{limit}
\end{eqnarray}
The $\Delta V$ required to evaluate  Eq. (\ref{cal0})
was set to $5\times10^{-3}~I_{c}$, $I_c$ being the
continuum intensity. This figure was estimated from the standard
deviation of $V$ as a function of $\lambda$
when $B_\mathrm{eff}$ becomes zero. Explicitly, 
the standard deviation of all
observed spectra was
represented versus $B_\mathrm{eff}$. Such scatter plot 
gives a non-zero Stokes $V$ standard deviation for $B_\mathrm{eff}=0$. This
value is used for $\Delta V$.

A second source of error affecting $B_\mathrm{eff}$ is due to the
contamination of the polarization signals with intensity.
Following the procedure described in Sect.~\ref{profiles},
spatially
restored spectra proportional to $I+V$ and $I-V$ are obtained for each 
point on the solar surface,
\begin{eqnarray}
S_1=& f_1\ (I+V)/2,\cr
S_2=& f_2\  (I-V)/2.
\label{eqmag0}
\end{eqnarray}
The circular polarization is estimated by subtracting them out (Eq.  
(\ref{profiles1})),
\begin{equation}
\widehat{V}=S_1- S_2.
\label{eqmag2}
\end{equation}
This would yield the true solar $V$ signal if $f_1 = f_2 = 1$. However,
due to the uncertainties of the reduction procedure (e.g., normalization
to the continuum intensity, insufficient flatfielding,  
non-linearities of the cameras, 
instrumental polarization induced by the telescope,
etc),
$f_1\not= f_2$ 
and $\widehat{V}$ is contaminated by the
intensity profile $I$,
\begin{equation}
\widehat{V}\simeq V+ {{f_1-f_2}\over{2}}\ I.
\label{eqmag3}
\end{equation}
Since $V \ll I$, 
the crosstalk with intensity  
seriously threatens the polarimetric accuracy of the 
measurements.
Equation (\ref{llest}) was devised to  automatically correct for this
crosstalk, at least to first order. One can estimate the
residual error due to crosstalk
with intensity as follows.  
Equations (\ref{llest}) and (\ref{eqmag3}) lead to
\begin{equation} 
\Delta B_\mathrm{eff}=B_\mathrm{eff}-B_\mathrm{eff0}={{f_1-f_2}\over 2} \Delta B,
\end{equation} 
with $B_\mathrm{eff0}$ the flux density to be retrieved if there were no
crosstalk, and $\Delta B$ characterizing the 
bias,
\begin{equation} 
\Delta B=\frac{\sum_j I(\lambda_j) C(\lambda_j)}{\sum_j C^{2}(\lambda_j)}.
	\label{i2v}
\end{equation} 
Note that $\Delta B$ only depends on the intensity.
We have estimated its value using a mean quiet Sun profile smoothed and 
sampled to mimic the observational procedures:
$¦\Delta B¦\sim\thinspace150$~G
for \lineone\ and $\sim\thinspace120$~G for  \linetwo. Keeping in mind that
the crosstalk $(f_1-f_2)/2$ has to be smaller than a few per cent (otherwise
the contamination would exceed the real Stokes $V$ signals),
the bias induced by the crosstalk with intensity $\Delta B\ (f_1-f_2)/2$
is at most a few G.
This figure is negligibly small compared to the
error produced by random noise (Eq.~(\ref{limit})).
Should we had used a different procedure to measure $B_\mathrm{eff}$,
the bias would be unbearable. For example, using only one wing 
of the line 
$¦\Delta B¦\sim 3.5\times 10^3$~G
for \lineone , which renders
$\Delta B_\mathrm{eff}\simeq 70$~G for
a 2\% crosstalk ($[f_1-f_2]/2=2\,\cdot\,10^{-2}$).
The huge difference with respect to our procedure
is due to the cancellation of positive and negative
contributions in Eq.~(\ref{i2v}) when the two 
wings are considered.

\subsection{Velocities}

A brief discussion on the velocities associated with the
magnetic signals is included in Sect.~\ref{spa_dis}.
These velocities are computed as the wavelength
of the minimum of a parabola that fits 
the core of the Stokes $I$ profiles.
The zero of the velocity scale is set by the mean
velocity across the full FOV, so it is affected by the convective
blueshift.

\subsection{Time series\label{time}}

We analyze the time variation of the magnetic signals
(Sect.~\ref{sectemp}). In order to produce a movie
from the individual speckle reconstructed magnetograms, the individual
snapshots of the series were
co-aligned allowing for a global shift
among them. Such  displacement is computed by
minimizing the difference between the successive 
broad-band images of the time series.
The mean broad-band image of the series
constructed in this way has a contrast
of the
order of 7\%. This figure has to be compared
to the contrast of the best image in the series,
which is some 12.5\%  (Sect. \ref{secsp}).

\section{Results\label{results}}

\subsection{Mean flux density of the magnetograms}
\begin{figure}
\resizebox{\hsize}{!}{\includegraphics{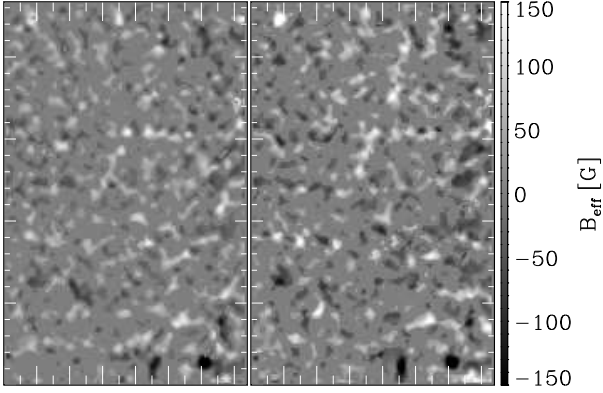}}
\caption{Magnetograms from  \linetwo\ (left) and \lineone\
(right). The signals below the noise level are set 
to zero. Dark and light represent different polarities,
as coded in the vertical bar. Tick marks are separated by 1 arcsec.
}
\label{mag} 
\end{figure} 
Figure \ref{mag} presents two simultaneous magnetograms taken in the two iron 
lines. (They correspond to the snapshot of the time series having
the best angular resolution,
which is the one that we analyze in detail.) 
They exhibit a {\em salt and pepper} pattern with 
patches of opposite polarity in close contact. 
Approximately {60\,\% of the FOV contains 
polarimetric signal above noise in one of the
spectral lines}. This large area coverage,
together with the fair magnetic sensitivity of the
measurements, yields
more magnetic flux than any previous observation of the 
photospheric IN magnetic fields (Sect.~\ref{other}). 
In order to quantify the
amount of flux, we have evaluated
the mean unsigned flux density in the observed region
$\overline{|B_\mathrm{eff}|}$,
\begin{equation}
\overline{|B_\mathrm{eff}|}\equiv \frac{1}{\tilde N} \sum_i |B_\mathrm{eff}^i|, 
	\label{bapp}
\end{equation}
where the sum includes only those pixels with signal above
noise in one of the spectral lines, and $\tilde N$ represents the total number of pixels
in the FOV. ($\overline{|B_\mathrm{eff}|}$ is the average $|B_\mathrm{eff}|$  over
the FOV once signals below noise are set to zero.)
The result of the computation yields
\begin{equation}
\overline{|B_\mathrm{eff}|}=\cases{21~{\rm G}, & for \lineone;\cr
	17~{\rm G}, & for \linetwo.}
	\label{meanfd}
\end{equation}
These mean flux densities are weakly affected by the noise of the individual 
$B_\mathrm{eff}^i$, since $\overline{|B_\mathrm{eff}|}$ results from the average of thousands of 
individual pixels. The fact that we average absolute values and
the use of a threshold do not modify this argument.
We carried out  Monte-Carlo simulations that show the
robustness of the estimates (\ref{meanfd}).
Using the full magnetograms (including
effective flux densities below the noise), we added random Gaussian noise according to 
our estimate of the noise level (Eq. (\ref{limit})).
The application of the definition (\ref{bapp}) to different realizations
of magnetograms with mock noise renders the same mean flux densities within 
a fraction of one G.

We have also computed the mean  {\em signed}
magnetic flux density for the magnetograms in Fig. \ref{mag} (with a definition similar
to the unsigned flux Eq. (\ref{bapp}) 
but using 
$B_\mathrm{eff}^i$ instead of $|B_\mathrm{eff}^i|$). It turns out to be,
\begin{equation}
\overline{B_\mathrm{eff}}=\cases{+2~{\rm G}, & for \lineone;\cr
	+3~{\rm G}, & for \linetwo.}
	\label{meanfd2}
\end{equation}
Note that $\overline{B_\mathrm{eff}}~\ll~\overline{|B_\mathrm{eff}|}$,
which probably constrains the physical mechanisms responsible
for the observed magnetic structure (they have to create large
unsigned magnetic fluxes and, simultaneously, low signed fluxes). 
This constraint is probably even more severe than the values inferred
from combining Eqs.~(\ref{meanfd}) and (\ref{meanfd2}), since
we cannot discard systematic errors 
of a few G affecting $\overline{B_\mathrm{eff}}$ (see, e.g., Sect.~\ref{magcal}).

%
\subsection{Magnetic field strengths and filling factors\label{mfs}}
The polarization signals obtained from the two spectral
lines  are correlated (see Fig. \ref{mag}). However,  
the magnitude of the effective flux density of \lineone, 
$B_\mathrm{eff}(6301)$, is systematically larger than the
effective field derived using \linetwo, $B_\mathrm{eff}(6302)$.
The difference is inferred from mean fluxes (e.g., 
from Eq. (\ref{meanfd})), since the noise prevents conclusions
based on individual measurements\footnote{The
	error of the ratio based on the observations
	in a single pixel
	is 
	$\Delta [B_\mathrm{eff}(6301)/B_\mathrm{eff}(6302)]
		\simeq 30\,\mathrm{G}/|B_\mathrm{eff}(6302)|$,
	which follows 
	from the law of propagation of errors, 
	Eq.~(\ref{limit}),
	and the ratio of effective
	fluxes that we will infer (Eq.~(\ref{ratio0})). 
	Since $|B_\mathrm{eff}(6302)|\le~150\,$G, then
	$\Delta [B_\mathrm{eff}(6301)/B_\mathrm{eff}(6302)]\ge 0.2$,
	which does not suffice to detect the 
	ratio~(\ref{ratio0}). One needs to average over many pixels to distinguish
	the ratio from unity.}.
We calculate the mean excess of $B_\mathrm{eff}(6301)$ with respect
to $B_\mathrm{eff}(6302)$ by means of a lest-squares fit that
accounts for the errors of both $B_\mathrm{eff}(6301)$ and
$B_\mathrm{eff}(6302)$. We only consider flux densities above noise.
Our best estimate is 
\begin{equation}
B_\mathrm{eff}(6301) /B_\mathrm{eff}(6302)\simeq 1.25\pm 0.14.
	\label{ratio0}
\end{equation}
The error bar corresponds to the standard 68\% confidence level, and
it is worked out in Appendix \ref{appb}, where we also reject
the hypothesis $B_\mathrm{eff}(6301)=B_\mathrm{eff}(6302)$ with
high confidence.
The 25\% systematic difference can be easily interpreted if
the intrinsic magnetic field strengths of  
the structures that give rise to the signals 
in the magnetograms exceeds 1 kG. 
Should the magnetic fields 
be intrinsically weak (say, a few hundred G),
the magnetograph equation holds
since the basic conditions for the approximation to be valid are satisfied.
First, the Zeeman
splittings of both \lineone\ and \linetwo\ are smaller 
than the Doppler width of the lines. Second, the magnetic
fields are dynamically weak, and they cannot modify
the thermodynamic conditions with respect to those
in the un-magnetized
atmosphere. Even if one does not spatially resolve the
magnetic structures, the calibration constants
derived from the observed Stokes $I$ are valid,
Eq.~(\ref{fill_factor}) applies,
and $B_{eff}(6301) / B_{eff}(6302)=1$.
Consequently, the fact that the observed ratio (\ref{ratio0})
differs from one implies the existence of strong fields. 
These qualitative arguments,
in the spirit
of the line-ratio method of \citet{ste73},
were already put forward
by \citet{soc02}.
The arguments can be made quantitative 
computing the ratio of effective field strengths
in different model atmospheres whose intrinsic field strengths
are known. We have carried out such calibration, and the
details are given in Appendix \ref{appa}.
The ratio of effective magnetic field
strengths versus the intrinsic field 
strength is computed in a variety of atmospheres
(Fig. \ref{calibration}). 
Two main results arise from this calibration:
\begin{itemize}
\item The observed ratio (\ref{ratio0}) indicates the existence
	of magnetic fields  larger than 1 kG
	in the photospheric layers where the observed
	polarization is formed. More precisely
	$900~{\rm G}~\le~B~\le~1600~{\rm G}$ when
	$1.1~\le~B_{eff}(6301)/B_{eff}~(6302)~\le~1.4$.
\item The fact that the ratio (\ref{ratio0}) corresponds
	to kG fields is almost insensitive to the details
	of the model atmosphere. This property  was expected
	according to the qualitative arguments given 
	above.
\end{itemize}

	Note that these conclusions do not discard the existence of sub-kG
magnetic field strengths in our data. Despite the fact that the {\em characteristic}
field strength seems to be kG, the scatter among the individual ratios
is very large. Consequently, 
there are many individual points whose ratio can be unity or smaller which,
according to the calibration above, implies sub-kG field strengths.

Structures with intrinsic kG magnetic field strength showing
20 G flux density have to
occupy only a small fraction of the solar
surface.
The simple 2-component model allows to estimate
the area coverage or filling factor.
The filling factor 
is just the ratio between effective and intrinsic
field strengths (Eq.~(\ref{fill_factor})).
If $B\sim$~1~kG then
\begin{equation}
\alpha\sim B_\mathrm{eff}/B\sim 20~{\rm G} / 1000~{\rm G}\sim 0.02.
\end{equation}
Only 2\% of the solar surface produces 
the observed signal. We observe them to cover 60\% of the
FOV because of the limited spatial resolution of the 
observations.
One can also employ the same order-of-magnitude calculation
to estimate the size $l$ of the magnetic concentrations.
If $M$ magnetic concentrations occupy a resolution element of size $L$,
then 
\begin{equation}
\alpha\sim (l/L)^2 M,
\end{equation}
which, together with Eq. (\ref{fill_factor}), renders
\begin{equation}
l\sim L \sqrt{B_\mathrm{eff}/(M\,B)}\sim\cases{75~\mathrm{km}& for M=1;\cr
					25~\mathrm{km}& for M=10;\cr
					10~\mathrm{km}& for M=50.}
		\label{sizes}
\end{equation}
We use $L=0\farcs5\equiv 363$~km,
	$B=1$~kG,
	and $B_\mathrm{eff}=40$ G,
the latter being the typical value of the signals above noise
in the magnetograms of Fig. (\ref{mag}).

\subsection{Spatial distribution and brightness
of elements with magnetic signal 
\label{spa_dis}}

\begin{figure}
\resizebox{\hsize}{!}{\includegraphics{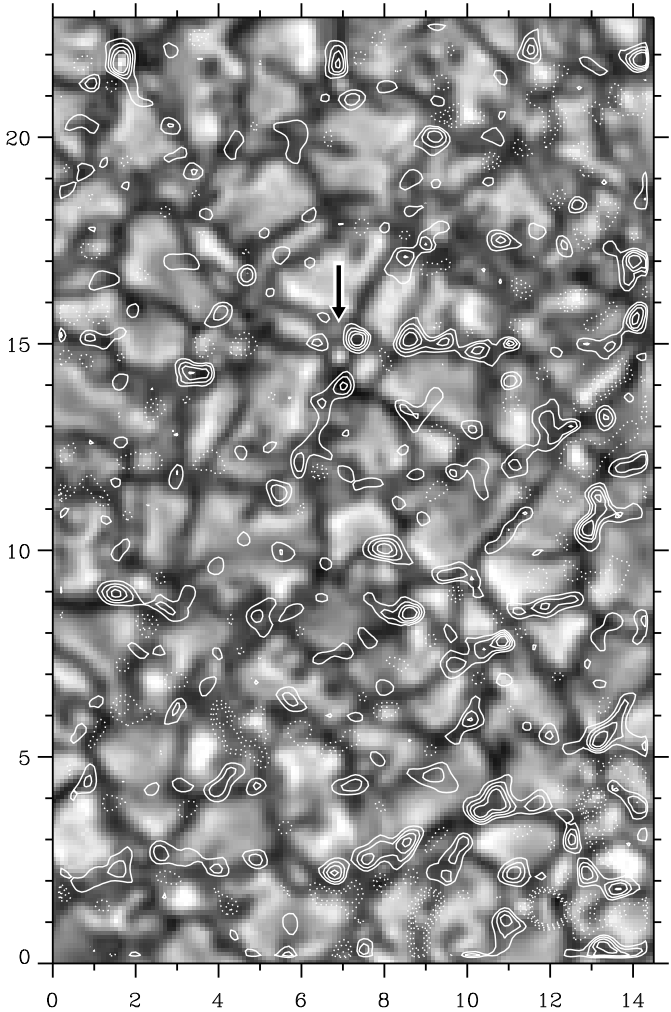}}
\caption{Speckle reconstructed broad-band image overlaid with the 
magnetogram  of
\linetwo\ with contours at $B_\mathrm{eff}=$~$\pm 30$, $\pm 50$,
$\pm 70$, and $\pm 90$~G. The solid and dotted contours indicate 
opposite polarities. The
distance between tick-marks is 1 arcsec, as indicated by the
labels of the axes. The arrow points out an isolated
bright point without associated polarization
signal.
}
\label{specmag} 
\end{figure}

\begin{figure} 
\resizebox{\hsize}{!}{\includegraphics{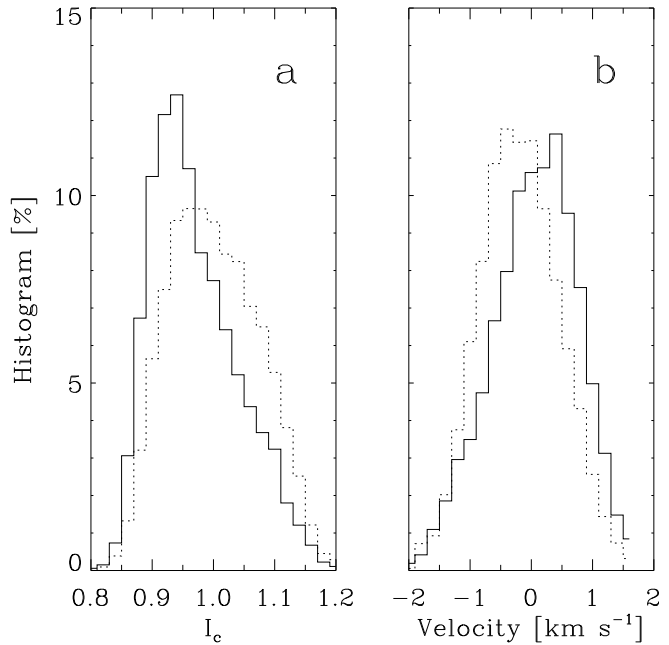}}
\caption{
        Histograms of continuum intensities ({\bf a}) and 
	velocities ({\bf b}) 
	for pixels with magnetic fields above the noise level
	(the solid lines) and below the noise level
	(the dotted lines). The intensities refer
	to the mean intensity, and the average velocity over the FOV 
        is set to zero. We adopt the 
	convention that redshifts are positive. The histograms correspond to \linetwo.
	}
\label{dist} 
\end{figure}

Figure \ref{specmag} shows the speckle reconstructed broad-band image 
and, overlaid to it, the  \linetwo\ magnetogram.  
Most of the magnetic fields 
are located in intergranular lanes. This tendency has been observed before 
\citep{lin99,lit02,soc03b} and is expected from 
numerical simulations of magneto-convection 
\citep[e.g.,][]{wei78,cat99a,vog03}.
However, some magnetic field
signals are also found in 
granules, in agreement with the observations of 
\citet{sto00} and \citet{kos01}.
This bias of the magnetic signals
towards intergranules is also evident from the inspection of
Fig.~\ref{dist}, which contains histograms of the 
distributions of continuum intensities and velocities.
The figure displays both, pixels with signals above noise
(the solid lines), and below noise (the
dotted lines). Some 65\% of the magnetic signals
of \linetwo\ are darker than 
the mean intensity, which indicates association
with intergranular
lanes. On the other hand, 60 \%  
of these pixels present redshifts, which 
correspond to downflows in the solar atmosphere and therefore trace intergranules.
The histograms based
on \lineone\ present similar trends.

\begin{figure}
\resizebox{\hsize}{!}{\includegraphics{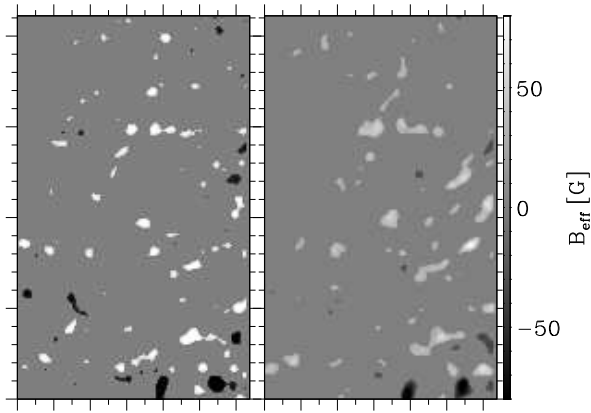}}
\caption{Left: magnetogram of the region showing 
	only strong signals, explicitly, flux densities  above 50\,G.
	The rest is set to zero.
	Note the regular pattern with a size similar
	to the 5-10\arcsec scale of the
	mesogranulation (tick-marks correspond to 1 arcsec).
	Right: mean magnetogram of the time series, which still shows 
	a pattern similar to that of the single snapshot. 
	In this case
	we set to zero those flux densities below 25\,G .
	}
	\label{mag2}
\end{figure}

The strong  signals in the 
magnetograms show a pattern whose characteristic size corresponds
to the mesogranular convective cells, i.e.,
with a spatial scale between 5 and 10 arcsec (Fig. \ref{mag2}).
Mesogranulation has been detected in
velocity \citep[e.g.,][]{nov81} and
intensity \citep[e.g.,][]{deu89}, and it is also 
recovered from numerical simulations of magneto-convection
\citep[e.g.,][]{cat01}. However, to the 
authors' best knowledge,
this is the first clear detection in polarization.
(For another observation suggestive of  
mesogranulation in polarized light,
see \citealt{tru03}, Fig.~4.)
Figure \ref{mag2}, left, portraits the magnetogram of the region
except that only fluxes above 50\,G have been represented.
One can guess a cellular pattern with a scale larger than
the granules but still smaller than the supergranulation 
(the latter with a scale larger than the FOV). 
This pattern is long-lived
since it shows up in the mean magnetogram of our
17 min time sequence  (Fig.~\ref{mag2}, right; see also
Sect.~\ref{sectemp}). Our finding should be regarded as an additional
observational proof for the existence of the mesogranular
convective scale.

\begin{figure}
\resizebox{\hsize}{!}{\includegraphics{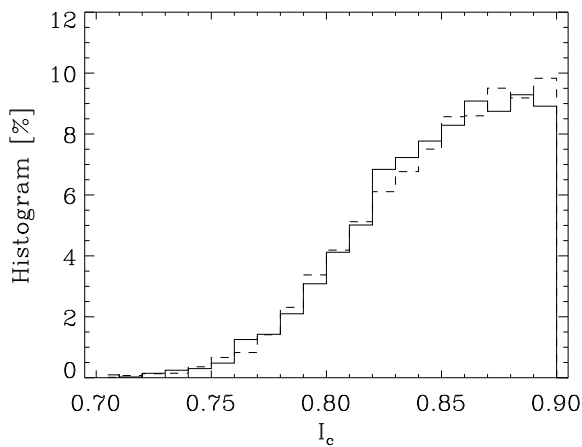}}
\caption{
Distribution of broad-band intensities in intergranules with and 
without magnetic field. Intergranules are selected as those points
with intensity smaller that 90\% of the
mean intensity. 
The two types of line distinguish the histograms obtained for 
points with magnetic field above the noise  (the solid line) and
those with no detected signals (the dashed line). There
is no obvious difference. The histograms 
provide the
percentage of points in each bin. 
}
\label{bright} 
\end{figure}

We have found magnetic structures
that typically harbor kG magnetic fields. According to the common
wisdom, they should be bright in broad-band images.
The presence of a kG field reduces the density of the
magnetized plasma and
thus lowers the opacity with respect to the ambient 
atmosphere.
One sees  deeper through 
the magnetized plasma
which (usually) means observing hotter layers that produce more light
(often called hot-wall effect; \citealt{spr76}). 
Except for a few cases (e.g., bright points at [1\farcs5, 22\arcsec],
[12\arcsec, 1\arcsec], or [8\arcsec, 10\arcsec] in Fig. \ref{specmag}),
we do not see  bright points associated with the
magnetic signals. Moreover,
the intergranules having magnetic signals
do not seem to be brighter than those without them
(see the histograms in
Fig. \ref{bright}).
This seeming inconsistency between the expected
but unobserved brightness 
may not be real. 
There are several ways to reconcile the observations
with the current paradigm, for example,
\begin{itemize}
\item the sizes of the individual kG magnetic concentrations (see
Eq.~(\ref{sizes}))
are small compared with the sizes of
intergranular lanes. Even very bright but small magnetic structures
may lead to dark signals when observed with the
kind of (good but still insufficient) angular resolution of the
broad-band image \citep[see][]{tit96}. 
\item If the temperature of the intergranules is low enough, 
	the expected reduction of opacity associated with the presence of
	kG magnetic fields may not produce a bright structure.
	The depression of the observed layers by a few tens of km
	may not be enough to reach 
	temperatures larger than those of the mean un-depressed photosphere
	\citep[see, e.g.,][ Fig.~14]{stei98}.
\item Due to a yet unknown physical process,
	there may be a tendency for the magnetic structures to 
	stay in the darkest intergranular lanes (e.g., they
	accumulate in the  strong downflows that tend to be particularly
	dark.) 
\end{itemize}

Note, finally, that the signals in Fig. \ref{specmag}
cannot be misidentified network magnetic 
concentrations, since they occur at a characteristic scale
much smaller than the supergranulation.

\subsection{Stokes $V$ asymmetries}

The Stokes $V$ profiles observed in quiet network and inter-network
regions show large deviations from the anti-symmetric shape
	\citep{san96,gro96,sig99,san00}. 
These so-called asymmetries carry meaningful information on the
structure of the magnetic atmosphere 
\citep[e.g.,][ and references therein]{san98c}.
It would have been desirable 
studying the asymmetries of our profiles in some detail. However,
the Stokes $I$ to $V$ contamination described in Sect.~\ref{magcal},
Eq. (\ref{eqmag3}), masks
the real  asymmetries in a significant way, and
we refrain from analyzing them. Once this caveat has been explicitly 
pointed out,
we also would like to mention that 
asymmetric Stokes $V$ profiles
are found everywhere in the FOV. In particular,
we observe Stokes $V$ profiles with only one lobe  and  profiles with
three lobes  
(see Fig. \ref{asymmetries}). 

\section{Temporal evolution
\label{sectemp}}

The duration (17 min) and cadence (50 s) of the series
of magnetograms are far from optimum to allow a detailed study
of the temporal evolution of the structures.
In addition, 
the polarization signals are close to the noise level,
and the seeing conditions vary along the sequence. These two
factors  induce  spurious time variations which complicate the
analysis. Despite these drawbacks, 
several conclusions on the time 
evolution of the magnetic signals can safely
be drawn. We extracted them from the inspection
of the sequence of magnetograms re-centered 
as described in Sect.~\ref{time}.

\begin{figure*}
\centering
\includegraphics*[width=17cm]{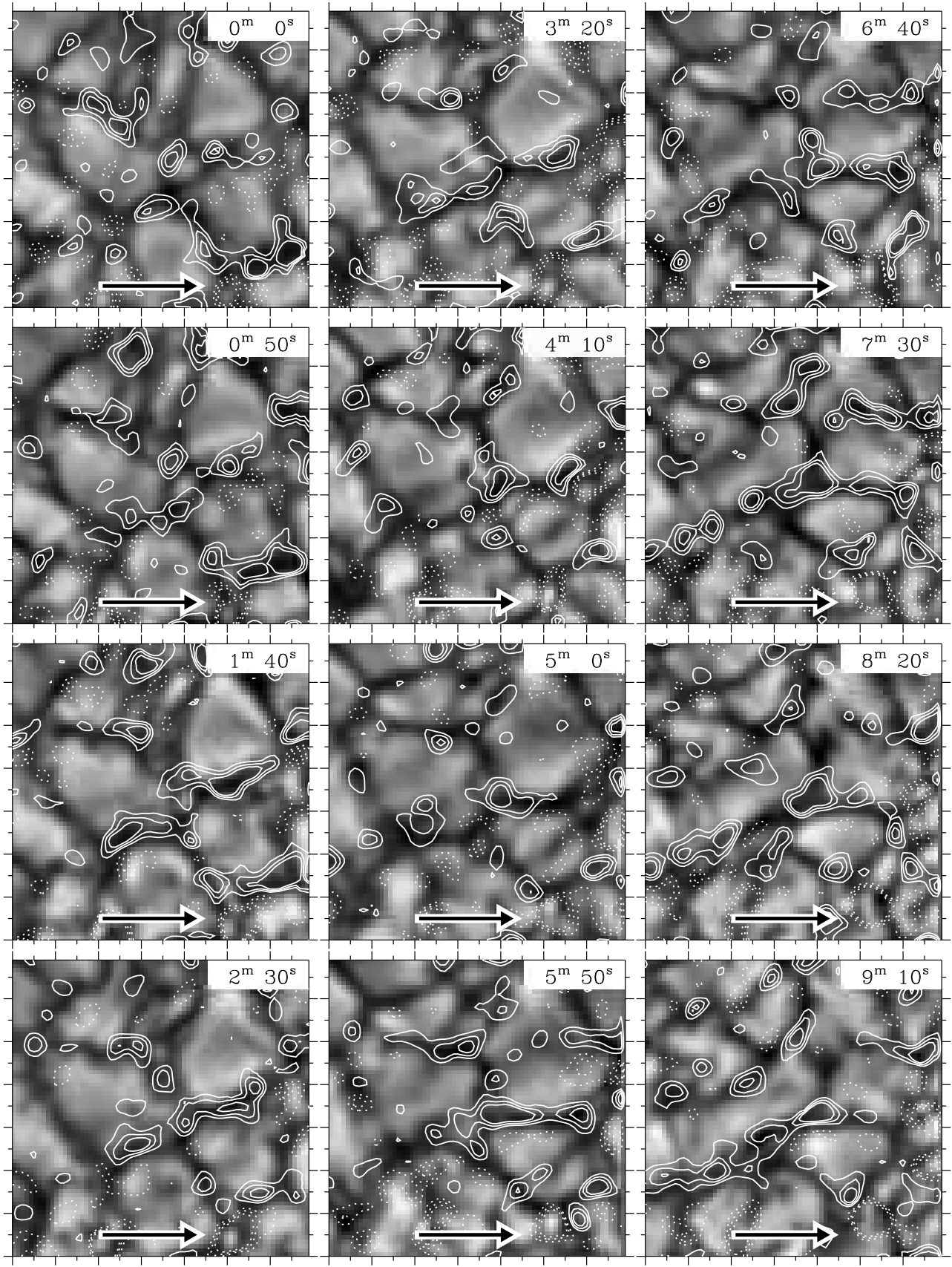}
\caption{
Temporal evolution.  Magnetograms overlapped to 
intensity images as in Fig.~\ref{specmag} but 
for a sub-field of 7\arcsec$\,\times\,7$\arcsec.
Time increases from top to  bottom and from left to right.
The different images are consecutive snapshots of the series
separated by 50 s (see the labels on the
images). The arrow points out a specific region
dominated by a magnetic concentration of negative
polarity
that interacts with positive polarity patches and, possibly, eats them away.
}
\label{mtemp} 
\end{figure*}

Many individual features maintain their identity
in successive time steps of the series
(see Fig. \ref{mtemp}, where time increases from top to
bottom and from left to right). In fact, 
some of the signals live longer than the total
time span, as it is proven by 
Fig. \ref{mag2}
where the strongest signals in the mean magnetogram of the series 
resemble those of any individual snapshot.
This mean magnetogram has been computed  as a 
weighted  average of the individual magnetograms of the series, with
the weights given by the inverse of the squared noise 
of the magnetograms (Eq.~\ref{limit}). 
In order to quantify the fraction of unsigned
flux that survives more than the time-span, we
have computed the
unsigned flux density for signals 
above noise level in the mean magnetograms. They
turn out to be,
\begin{equation}
\overline{|B_\mathrm{eff}|}=\cases{13~{\rm G}, & for \lineone;\cr
	9~{\rm G}, & for \linetwo,}
	\label{meanmagfd}
\end{equation}
They represent  60\% of the flux density in the best individual
magnetogram (Eq. (\ref{meanfd})).
It is tempting to interpret this figure as the
fraction of IN magnetic structures that live longer
than 17 min. However, the interpretation
is not so straightforward since we detect polarization
signals rather than magnetic structures.
Seemingly long-lived signals may not necessarily
correspond to stable magnetic structures. 
Numerical simulations of quiet Sun magneto-convection
show persistent downdrafts that continuously gather 
magnetized plasma in specific locations 
\citep[e.g.][]{cat99a,emo01}. The 
plasma sinks down and disappears  along the
downdrafts, and it is continuously replaced by new plasma.  
The system of persistent downdrafts outlines  
mesogranular cells, which live much longer than the 
individual magnetic structures.
This theoretical scenario is consistent with the
observed magnetograms and, in particular, with the
existence of a long lasting polarization pattern
that we have already associated with the
mesogranulation
(see Sect.~\ref{spa_dis}). Consequently, the existence
of signals in the mean magnetogram is
interpreted here as a proof of the temporal coherence
of the mesogranular pattern, rather than the persistence
of individual magnetic concentrations.

	On top of the rather stable pattern described above,
large variations of the magnetic signals occur
between snapshots (Fig.~\ref{mtemp}).
Some polarization
signals migrate following the granular motions to
remain in the intergranular spaces. There are places
where opposite polarities approach each other 
and seem to (partly) cancel out; the  
arrow in Fig. \ref{mtemp} indicates a negative polarity
patch (dotted line) that interacts, and possible eats away,
neighboring positive polarity contours (compare the snapshots
at 0$^\mathrm{m}\,0^\mathrm{s}$ and
4$^\mathrm{m}\,10^\mathrm{s}$).
More often signals grow and fade with no obvious interference 
with  patches of opposite polarity. 
As we pointed out above,
these variations only trace changes in the polarization
signals, which do not necessarily imply the evolution of the
magnetized plasma. Actually, one may have large 
variations of polarization 
without substantial changes of the underlying magnetic structure.
For example,
\citet{san00c} describes how 
the polarization of kG magnetic concentrations can 
vanish
upon a small increase of the field strength. This effect is
mentioned here because it
may be responsible for the
lack of magnetic signal associated to 
the bright point indicated by an arrow in Fig. \ref{specmag}.
Subsequent magnetograms of the series show that this bright point
is co-spatial with polarization signals.

The mean unsigned flux density is only slightly different for the different
snapshots of the series. The 
standard deviation of these changes is some 10\% for
\linetwo , whereas it reaches 20\% for \lineone . 
There is a tendency 
for the snapshots of largest continuum contrast to have
the largest flux densities.

\section{Consistency of the results\label{consistency}}
The unsigned flux densities that we find  are larger 
than the values found 
hitherto. 
The purpose of this section is to spell out  
arguments that support the 
reliability of our results. 
The magnetograms have self-consistency
since independent parts of the data set provide similar
results. They are also consistent with previous
observations having different sensitivities
and angular resolutions, once these differences
are properly taken into account. Finally, the signals
meet several theoretical prejudices which are
independent of the observation. In particular,
the signals appear in the intergranular lanes
and show a conspicuous and stable 
mesogranular pattern.

\subsection{Self-consistency}
We have a time sequence of 
magnetograms taken in two different spectral lines.
Magnetograms differing in time or spectral line are
independent, and they can be analyzed independently
to check the self-consistency of various results.
We have identified the following properties of the data set
that support the internal agreement between independent subsets,
\begin{itemize}
\item the simultaneous magnetograms obtained from the two lines
	are very similar (Fig. \ref{mag}).
\item Many patches in the magnetograms are larger than the
	resolution of the observation, implying a spatial
	coherence of the signals difficult to ascribe to noise.
\item The signals are co-spatial with the
	dark intergranular lanes (Sect.~\ref{spa_dis}).
	Again,
	this relationship is difficult to explain 
	as due to  noise. 
	One might suspect that the
	misalignment of the two circularly
	polarized beams subtracted to determine Stokes $V$
	creates false polarization signals. This contamination
	would be maximum where the spatial gradients in 
	intensity or velocity are maximum \citep[e.g.,][]{lit87}.
	The maximum gradients occur in the transitions between
	granules and intergranules, which is not where the detected
	signals are. We have checked that
	the scatter plots of polarization signals versus
	spatial gradients show no obvious correlation.
\item Many polarization patches maintain their identity 
	in several successive magnetograms of the series.
	In particular, the mean magnetogram retains 
	some 60\,\% of the peak flux density (Eq.~(\ref{meanmagfd})),
	a fact once
	again difficult to explain as due to noise. 
	Should the signals be noise, the mean magnetogram
	would show signals much smaller than that
	of individual frames (i.e., only $\sim 1/\sqrt{20}\sim20$\,\% 
	of the individual signals).
\end{itemize}

\subsection{Comparison with other measurements\label{other}}
Our observations differ in angular resolution and
sensitivity from previous measurements. These factors
have to be taken into account before they can
be properly compared.
We have to cut down the resolution of the
magnetogram and then change the sensitivity of the
measurement.

\begin{figure}
\resizebox{\hsize}{!}{\includegraphics{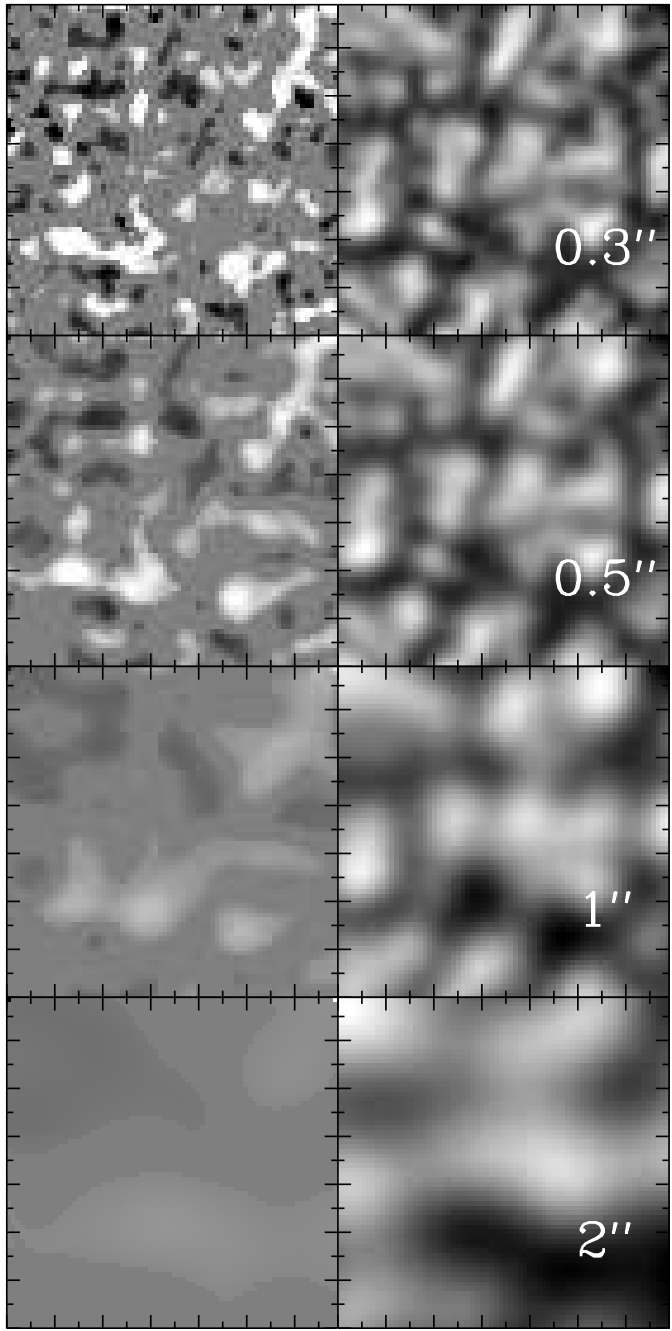}}
\caption{Magnetograms (left) and continuum intensities (right) for
different seeing conditions as indicated by the insets. All
magnetograms have the
same scale, which saturates at -60 G and 60 G. The field of
view is only 7\arcsec$\,\times\,$7\arcsec.
}
\label{magseeing}
\end{figure}

\begin{figure}
\resizebox{\hsize}{!}{\includegraphics{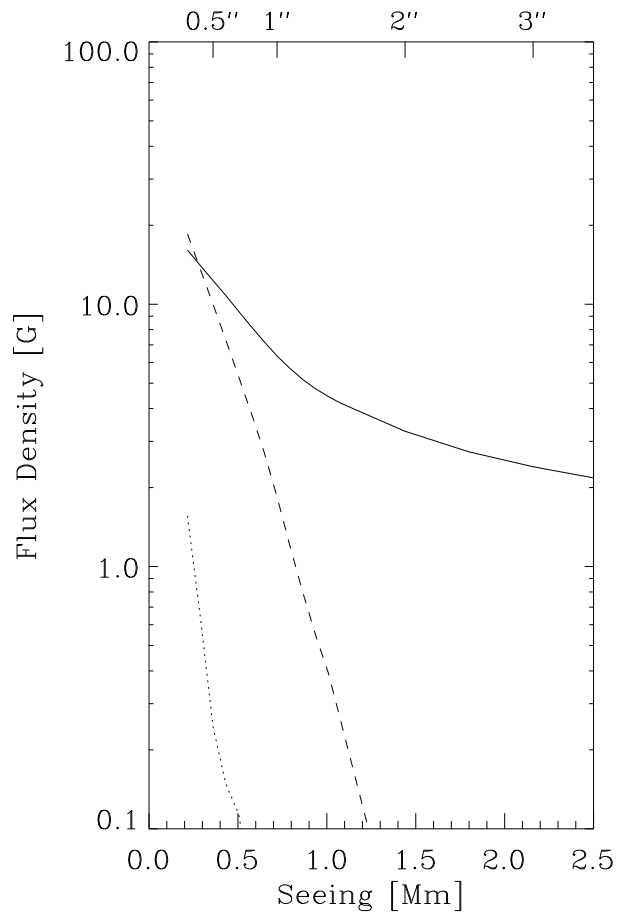}}
\caption{Mean unsigned \linetwo~flux density 
that remains in the magnetograms when
observed with the seeing given in the abscissa.
If all signals above noise are considered, one
is left with the solid line. If only
signals above 20\,G are considered, then the
reduction 
is much more severe as represented by the 
dashed line. Finally, when very large flux densities
are considered (signals larger than 100\,G) 
almost no signal remains (the dotted line).
}
\label{fluxseeing}
\end{figure}

	The speckle reconstructed images 
can be regarded as the {\em true} solar image down to the
cutoff frequency. In the case of our magnetograms,
the cutoff corresponds to a period
of 0\farcs5 (see
Sect.~\ref{secsp}, Fig. \ref{pow}). We model the effect of seeing on the
measured images by convolving the speckle reconstructed
images with a Gaussian point spread function. 
This smoothing is carried out independently
for each single wavelength.
Then the different wavelengths are combined and
processed to get magnetic flux densities
in the same way as the 
original images.
The  FWHM of the Gaussian kernel is used to quantify the seeing.
Figure \ref{magseeing} shows how the 
magnetograms (left) and
broad-band 
intensities (right) 
appear under
various seeing conditions\footnote{One
might regard as inconsistent modeling a 0\farcs3
seeing when the cutoff frequency of the observation is only 
0\farcs5. Note, however, that the speckle reconstructed
images contain all the power down to the
cutoff, and this power is partly  reduced
under any seeing condition, including seeing
with FWHM smaller than the cutoff.}:
0\farcs3,
0\farcs5, 1\arcsec~ and
2\arcsec . The field of view of the section 
of the magnetogram that we present is only
7\arcsec$\,\times\,$7\arcsec.
Even when the seeing is fair (1\arcsec), only
traces of the signals in the speckle 
reconstructed images are left. 
The unsigned flux density that survives a 
given seeing is shown  in Fig. \ref{fluxseeing}.
It contains the signals above three different levels 
of noise (or different sensitivities): 
100\,G, 
20\,G and the actual noise in 
the magnetograms.
The decrease of angular resolution reduces the signals 
due to cancellation between close 
opposite polarities. For an angular resolution
corresponding to 1\arcsec ($\equiv 0.75$\,Mm),
the mean unsigned flux density  become of the order of 7 G.
This value agrees with previous determinations 
based on scanning spectro-polarimeters that reach 1\arcsec\ resolution 
(\citealt{san00} and \citealt{lit02} obtain some 10\,G).
\citet{san03} collect different values from the literature,
showing a magnitude and  trend
that resembles
the solid line in Fig. \ref{fluxseeing}.
In particular, \citet{wan95}
point out a flux density of 1.65\,G for a 2\arcsec\ resolution, which
is not far from the value that remains in our magnetograms
for this resolution (some 2.5\,G; Fig. \ref{fluxseeing}).
On the other hand, magnetograms with an angular resolution
similar to those presented here exist in the
the literature
\citep{kel95,kos01,ber01}.
However, they have a noise level
between 50\,G and 150\,G, which
exceeds the limit required to detect most of the
signals in our  magnetograms. This lower sensitivity
can easily explain why the IN signals that we detect
have been missed so far. Figure \ref{fluxseeing} shows that
the mean flux density drops below 1-2\,G when 
only signals above 100\,G are considered.
In short, our magnetograms are consistent with previous
observations when the limitations of angular resolution
and sensitivity are taken into account.

We find imbalance between the two polarities so that
there is a net magnetic flux across the region (Eq.~(\ref{meanfd2})). 
We do not claim that the effect
is real since it is at the level of systematic
effects that may be important (see Sect.~\ref{magcal}).
However, imbalances of similar magnitude have been found in the
IN by other authors \citep[see][]{lit02},
and consequently, the imbalance may be true. If so, it would be
one more aspect of our observation being consistent with
previous measurements of IN field properties.

\section{Conclusions\label{conclusions}}

We have obtained a time sequence 
of speckle reconstructed 
quiet Sun magnetograms showing magnetic signals
that cover  60\% of an Intra-Network region
(Fig.~\ref{specmag}).
The signals appear as patches of opposite polarity 
often located close to each other.
The structures are not evenly spread out,
but they tend to accumulate in the intergranular
lanes. Those  signals of large
flux density trace a network whose 
scale is larger than the granules and
smaller that the supergranules. We associate
this network with the mesogranulation,
difficult to observe in velocity and intensity 
but very conspicuous in polarization
(Fig.~\ref{mag2}). The mesogranular
pattern lives longer than
the time span of the series (17 min). 

The effective flux density 
deduced from \lineone\ is systematically larger
than the flux density of \linetwo\ 
(Eq.~(\ref{ratio0})). We interpret
this difference as due to the presence of kG field strengths
in  the underlying magnetic structures.
These field strengths 
saturate, differently, the polarization signals
of the two spectral lines.
The inference of kG, however, does not rule out that 
weaker field strengths are responsible for
some of the observed signals (see Sect.~\ref{mfs}).
Since the flux density  is proportional to
both the field strength and the filling factor,
we are forced to conclude that the true magnetic 
concentrations producing
the polarization cover only
a small fraction of the surface; some 2\%.

The magnetic signals that we observe
meet several theoretical prejudices:
they appear in intergranular lanes, trace 
mesogranulation, vary in short timescales, and have
complex topology with opposite polarities in contact.
These properties are easy to interpret
as the result of the interaction between 
granular convection and magnetic fields,
no matter whether the fields are produced by a 
local dynamo \citep{pet93,
	cat99a,
	emo01}, 
	a global dynamo \citep{ste02},
        or they represent
	re-processed materials
	left by old active regions
	\citep[e.g.,][]{sch03,vog03}.
There are other observational properties which do not 
accommodate so easily within the existing paradigms.
The field strengths are typically kG, so a process has
to concentrate the fields to this
level. The convective collapse
\citep{par78,spr79}, 
devised for network magnetic concentrations,
may have difficulty to concentrate the low magnetic flux
features that we detect \citep[see][]{san01c}. Yet another
observational constraint difficult to satisfy 
is the large difference between the signed and
unsigned flux densities. Whatever 
physical mechanism generates the observed fields,
it has to produce a distribution of magnetic field whose
mean flux is 
10 times smaller than the local
fluctuations of flux. 

\citet{san03} compile
values for the mean unsigned flux density of the IN fields
measured by different authors.
They show a tendency to increase as the angular resolution improves,
but none of them exceeds 10\,G. 
Our magnetograms contain more magnetic flux than the
values reported hitherto.
We detect more polarization signals  
probably due to the unique combination of high
spatial resolution and good sensitivity.
The magnetic flux detected in the IN 
critically depends on the polarimetric
sensitivity and the angular resolution. 
If these factors are properly taken into account, 
our estimates of unsigned flux density 
are compatible with the previous measurement.

We find evidence that our angular resolution and
sensitivity do not suffice to detect all the magnetic flux
existing in the quiet Sun. The flux density in the
magnetograms drops off
upon artificial reduction of 
polarimetric sensitivity or spatial resolution
(Fig.~\ref{fluxseeing}). The extrapolation  
of this trend predicts 
a notable increase of signals 
upon improvement of the observational limitations.
On the other hand, we have concluded that the observed signals
are tracing kG magnetic fields. 
There are sensible theoretical arguments pointing out
that the spectral lines used to measure are biased 
towards these kG fields
	\citep{san00,soc03}.
In other words, 
magnetic concentrations of sub-kG field strength could 
have been 
easily overlooked by our magnetograms. 
These and other reasons suggest that the 
IN flux detected so far 
should be regarded as a lower limit
to the true flux. Thus any effort to
improve the sensitivity and/or angular resolution 
is likely to be rewarded with new magnetic flux.
 

\begin{acknowledgements}
Thanks are  due to K. Jan\ss en for letting us use her image reconstruction
routines.
She, as well as J. Hirzberger and O. Okunev, helped us during the observations.
M. Sch\"ussler prompted us to search for
	mesogranulation in the magnetograms.
IDC acknowledges support by the Deutsche Forschungsgemeinschaft (DFG) through
grant 418\,SPA-112/14/01. 
The Vacuum Tower Telescope is operated by the Kiepenheuer-Institut f\"ur
Sonnenphysik, Freiburg, at the Spanish Observatorio del Teide of the
Instituto de Astrof\'\i sica de Canarias.
%
The work was partly supported by the Spanish {\em Ministerio de Ciencia
y Tecnolog\'\i a} and FEDER,
project AYA2001-1649.
\end{acknowledgements}


\def\josa{JOSA}

\appendix 
\section{Ratio $B_{\mathrm{eff}}(6301)/B_{\mathrm{eff}}(6302)$ \label{appb}}

We want to estimate the typical ratio of effective magnetic fluxes
taking into account that (a) the measured magnetic fluxes of the 
two lines have errors, and (b) these errors 
are not negligible.
We invoke several results from statistics, which
can be found in standard textbooks \citep[e.g.,][]{mar71}.
For the sake of conciseness, and  only in this
Appendix, the following notation is employed
\begin{eqnarray}
B_{1i}\equiv& B_{\mathrm{eff}}(6301){\mathrm{~in~the{\it ~i-th}~pixel}},\cr
B_{2i}\equiv& B_{\mathrm{eff}}(6302){\mathrm{~in~the{\it ~i-th}~pixel}}.
\end{eqnarray}

The fluxes observed in each pixel have
true values, $B_{1i}^0$ and $B_{2i}^0$,  plus noise, 
$\Delta B_{1i}$ and $\Delta B_{2i}$,
\begin{eqnarray}
B_{1i}=&B_{1i}^0+\Delta B_{1i},\cr
B_{2i}=&B_{2i}^0+\Delta B_{2i}.
	\label{eq_appb_1}
\end{eqnarray}
We assume the noise of the two fluxes to be
independent,
since they
come from separate data. We also assume 
a linear relationship
between the true fluxes,
\begin{equation}
B^0_{1i}=m\, B_{2i}^0,
	\label{linear}
\end{equation}
being  $m$ the ratio that we want to estimate.
The above expression, combined with Eq. (\ref{eq_appb_1}), renders
a linear relationship
between the two observables $B_{1i}$ and $B_{2i}$ that
explicitly includes noise,
\begin{equation}
B_{1i}=m B_{2i}-m \Delta B_{2i}+\Delta B_{1i}.
\end{equation}
As usual, $m$ is determined by means of a least-squares fit.
We
chose to minimize the merit function
\begin{equation}
{\rm X}^2(m)=\Large\sum_{i=1}^N
	{{(B_{1i}-mB_{2i})^2}\over{\sigma^2_{1i}+m^2\sigma_{2i}^2}},
	\label{eq_appb_2}
\end{equation}
where $\sigma^2_{1i}$ and $\sigma_{2i}^2$ are the variances of the 
probability density functions describing the noise 
$\Delta B_{1i}$ and $\Delta B_{2i}$,
\begin{eqnarray}
\sigma^2_{1i}\equiv&V\{\Delta B_{1i}\}=V\{B_{1i}\},\cr
\sigma^2_{2i}\equiv&V\{\Delta B_{2i}\}=V\{B_{2i}\},
\end{eqnarray}
whose expectation values are assumed to be zero,
\begin{equation}
E\{\Delta B_{1i}\}=
E\{\Delta B_{2i}\}=0.
\end{equation}
(The symbol $V\{~\}$ stands for the variance of a random
variable, and it should not be confused with the Stokes $V$
parameter.) 
$N$ in Eq. (\ref{eq_appb_2}) represents the number of pixels  used in the
estimate.
This particular merit function has been selected because
it presents several practical advantages:
\begin{itemize}
\item $E\{B_{1i}-mB_{2i}\}=0$, $V\{B_{1i}-mB_{2i}\}=\sigma^2_{1i}+m^2\sigma_{2i}^2$,
and so, invoking the central limit theorem,
each term of the definition (\ref{eq_appb_2}) is the
square of a random variable distributed according to 
a n(0,1) (i.e., a normal
distribution with mean of zero and  variance of one).
Should the noise of different pixels be independent,
the random variable ${\rm X}^2$ follows a
$\chi^2$ distribution, which makes it
easy setting confidence intervals and performing
statistical tests (see below).
\item The ratio of flux densities derived from the fit does not depend on whether
	we estimate $m$ or $m^{-1}$, i.e., on whether we start up from
	the relationship
	(\ref{linear}) or its inverse
	$B^0_{2i}=m^\prime B_{1i}^0$. In this second case one
	would obtain $m^\prime=m^{-1}$.
\item ${\rm X}^2$ minimizes the relative errors of the fit  
including both the errors of abscissae and ordinates.
\end{itemize}

\begin{figure}
\resizebox{\hsize}{!}{\includegraphics{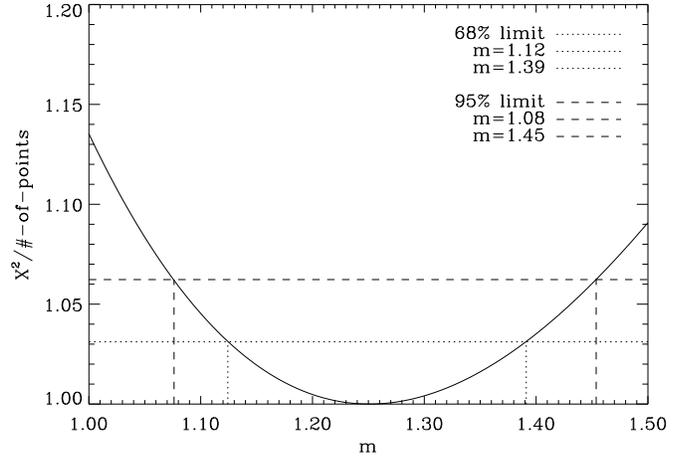}}
\caption{
	Merit function minimized to estimate the ratio 
	$B_{\mathrm{eff}}(6301)/B_{\mathrm{eff}}(6302)$. This 
	ratio is represented
	in abscissae using the symbol $m$.
	The merit function $X^2$ has been normalized to the number 
	of points used in the estimate, and it has a minimum
	around $m\simeq 1.25$. The horizontal and
	vertical lines give the confidence intervals for two
	significance levels, as indicated by the inset.
}
\label{chi2}
\end{figure}

We carry out the non-linear ${\rm X}^2$ minimization  by brute force, 
representing ${\rm X}^2$ versus $m$ and selecting the extreme. Figure \ref{chi2}
shows ${\rm X}^2(m)$ when $\sigma_{1i}$ and $\sigma_{2i}$
are the errors
of the fluxes estimated according to Eq. (\ref{limit}). 
If these variances were exact, then the expected value of 
each term of Eq. (\ref{eq_appb_2}) is one and
\begin{equation}
E\{{\rm X}^2\}=N.
	\label{expected}
\end{equation}
When the original error estimates are used,
the value of ${\rm X}^2$ at the minimum
becomes slightly below this expected value
(\ref{expected}), which we interpret as 
an overestimate of the variances
when using Eq. (\ref{limit}). 
We cure the
excess decreasing all the errors by 10\% to force
that ${\rm X}^2$ equals N at the minimum. 
Figure (\ref{chi2}) shows that such minimum of ${\rm X}^2$
occurs when
\begin{equation}
m=1.25\pm{0.14},
	\label{value_m}
\end{equation}
where the error bar represents the standard 68\% confidence interval.
In order to evaluate this  confidence interval, one must
know the probability density function that describes 
the merit function ${\rm X}^2$ considered as a random variable.
If all the points of the magnetograms were independent,
then  ${\rm X}^2$ would be distributed according to 
a $\chi^2$ distribution with $N$ degrees of freedom,
implying
\begin{equation}
V\{{\rm X}^2\}=2N.
	\label{expected2}
\end{equation}
However, we smooth the original magnetograms with a
$5\times 5$ pixels boxcar (Sect.~\ref{secsp}), 
so the different pixels are not independent.
This smearing reduces the degrees of freedom of 
${\rm X}^2$ and increases its variance with respect to
that in Eq.~(\ref{expected2}). 
If the magnetograms are divided
in groups of $5\times 5$ neighboring pixels, the
central points of all the groups are independent.
Using only these central pixels to construct
the merit function, then the new merit function
${\rm X}^2_{\xi}$ follows a
$\chi^2$ distribution  with $N/25$ degrees of freedom,
since now the sum (\ref{eq_appb_2}) contains $N/25$ summands.
If all the members of each group connected by the 
smoothing were identical,
then ${\rm X}^2=25\,{\rm X}^2_{\xi}$, and so
${\rm X}^2$ would have a variance 25 times larger than 
that in Eq. (\ref{expected2}),
\begin{equation}
V\{{\rm X}^2\}=V\{25\,{\rm X}^2_{\xi}\}=
	25^2\cdot 2N/25=25\cdot 2N.
	\label{expected3}
\end{equation}
Equations (\ref{expected2}) and (\ref{expected3}) represent
the two extreme limits corresponding to the cases
where all 
pixels which are not independent  due to the smoothing
are identical (Eq.~(\ref{expected3})), and when
all the pixels are independent (Eq.~(\ref{expected2}). In
general 
\begin{equation}
V\{{\rm X}^2\}=a^2\,2N,
\end{equation}
with $a=1$ if all the pixels have independent  noise
and $a=5$ if 
pixels linked by smoothing have identical noise.
In our problem $a$ has to lie in between these two extrema.
We estimate its value 
carrying out a Monte Carlo simulation. We produce
2D maps of n(0,5) random variables with the size
of our magnetograms. These maps are smoothed with a
$5\times 5$ pixels boxcar, so that each individual
point of the smoothed maps is a n(0,1) 
random variable (i.e., 
the same kind of {\em dependent} random variables 
that contribute to ${\rm X}^2$).
Then we randomly select $N$ pixels of the map
to compute ${\rm X}^2$ according to
the definition (\ref{eq_appb_2}).
The standard deviation
of ${\rm X}^2$ among several
hundreds independent realizations of the numerical experiment
renders
\begin{equation}
a\simeq 2.
	\label{value_a}
\end{equation}
In this intermediate case  ${\rm X}^2$ does not
necessarily follow a $\chi^2$ distribution,
however, invoking
the central limit theorem, one can argue that
the random variable defined as
\begin{equation}
Y=({\rm X}^2-N)/(a\sqrt{2N}),
	\label{master_eq}
\end{equation}
follows a n(0,1). This argument, together with the value of $a$ given
in (\ref{value_a}), let us set confidence intervals. For example,
the standard 68\% confidence interval, corresponding to $|Y| <1$, yields
the values of $m$ included in Eq. (\ref{value_m}). They
are computed as those $m$ with
${\rm X}^2(m)=N+ a\sqrt{2N}$ (see
Fig. \ref{chi2}).

Equation (\ref{master_eq}) can also be used to 
reject the hypothesis $m=1$ with high confidence
(i.e., to reject that the effective flux densities of the two
lines are identical). Assume that $m=1$ and that
the error estimates are correct.
The high value of ${\rm X}^2(1)/N$ 
in Fig. \ref{chi2} may be
produced by an unusual fluctuation of the
random variable in our particular realization. 
Using Eq. (\ref{master_eq}), the
probability of having a ${\rm X}^2$ equal or larger than the
value that we measure is smaller than
$10^{-4}$, since $[{\rm X}^2(1)-N]/a\,[2N]^{1/2}\sim 4$.
In other words, 
the two lines show different magnetic flux densities
with 99.99\% confidence.

The estimate (\ref{value_m}) uses those pixels in the
magnetograms with signal above noise.
As we have mentioned,
$\sigma_{i1}$ and $\sigma_{i2}$ come from
the values obtained
from the individual estimates of noise described 
in Sect.~\ref{magcal}. However, we
tried other possibilities: (a) using
all pixels including those with signals below noise, 
(b) using $\sigma_{i1}$ and $\sigma_{i2}$ constant and
given
by the mean values obtained in Sect.~\ref{magcal}, (c) forcing
$\sigma_{i1}=\sigma_{i2}$, and (d) adopting 
$a=3.5$. In all these other cases
$m\sim 1.2$ and the chance 
$m=1$ turns out to be  improbable.

\section{Calibration of
the magnetic field strength determination\label{appa}}

We want to calibrate the relationship between the ratio 
of flux densities derived from \lineone\ and \linetwo ,
$B_{eff}(6301) / B_{eff}(6302)$, and the intrinsic field strength.
First, Stokes $I$ and $V$ profiles are synthesized
in a set of model atmospheres of known field strength.
Second, the synthetic profiles are smeared out with a 
band-pass filter similar to the PFI used in the observations.
Third, the profiles are sampled in wavelength according to the
observed profiles. Finally, the definition of
$B_\mathrm{eff}$ is applied (Eq.~(\ref{llest})).
\begin{figure}[hb]
\resizebox{\hsize}{!}{\includegraphics{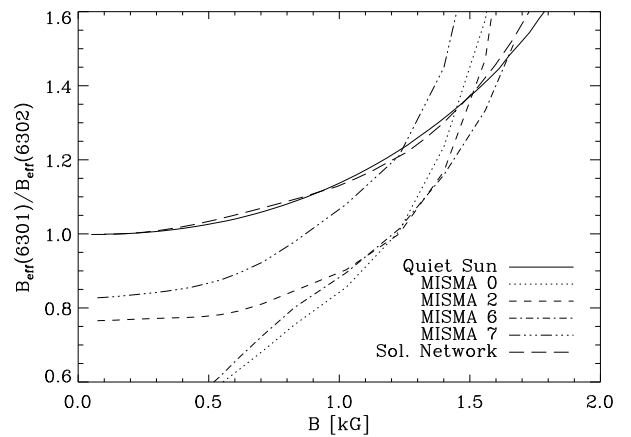}}
\caption{
Ratio of effective magnetic fields versus true magnetic
field strength in different  model atmospheres. The observed ratio
(Eq.~\ref{ratio0}])
agrees with the 
modeled values only if the intrinsic fields are larger than, say,
1 kG. This result is independent of the details of the model
atmosphere. Different types of line represent different model 
atmospheres, as coded in the inset.
}
\label{calibration}
\end{figure}

\begin{figure*}
\sidecaption
\includegraphics[width=12cm]{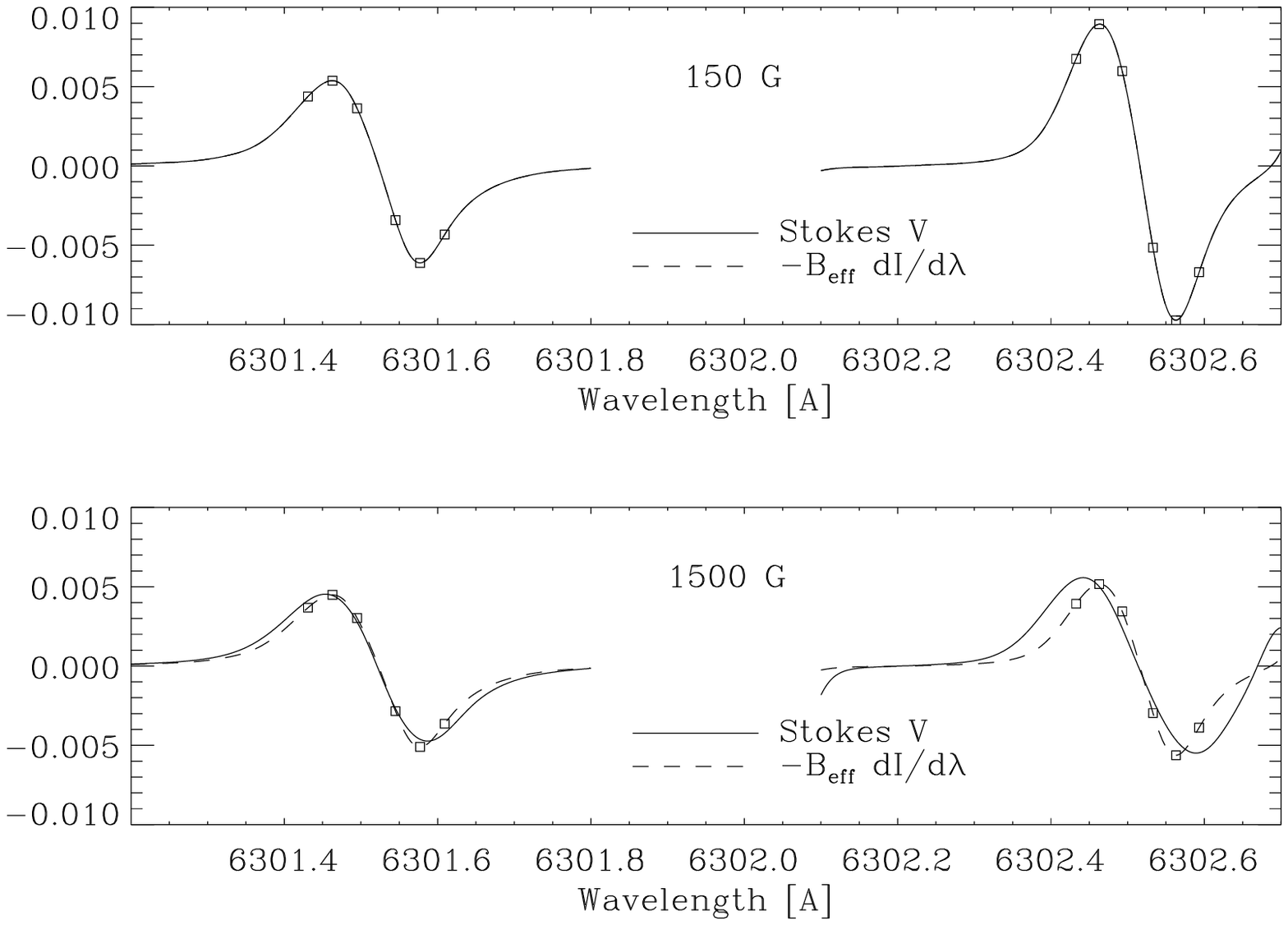}
\caption{Solid lines: synthetic Stokes $V$ profiles of \lineone\ and
\linetwo\  when the magnetic field strength is 150 G (top)
and 1500 G (bottom). The figures also include 
the derivative of the corresponding Stokes $I$ profiles,
scaled so as to  best-fit the Stokes $V$ (the dashed lines).
The square symbols point out the six wavelengths used 
to determine
the effective magnetic field strength $B_\mathrm{eff}$ ($\sim 30$\,G in both
cases).
}
\label{example}
\end{figure*}

Synthetic ratios are presented in Fig.~\ref{calibration}.
Three different types of model atmospheres are considered.
The first one assumes 
a magnetic atmosphere with
thermodynamic properties 
identical to that of the
quiet Sun. Then, under the assumption of longitudinal magnetic 
field\footnote{This assumption could be relaxed since it 
is not critical.},
the polarization of such atmospheres is
\begin{eqnarray}
V_m(\lambda)=&\frac{1}{2}\big[I_q(\lambda+\lambda_B)-I_q(\lambda-\lambda_B)\big],\cr
I_m(\lambda)=&\frac{1}{2}\big[I_q(\lambda+\lambda_B)+I_q(\lambda-\lambda_B)\big],
\end{eqnarray}
where $I_q$ represents the Stokes $I$ line profile 
produced by the unmagnetized quiet
Sun. The symbol $\lambda_B$ stands for the Zeeman splitting of the 
line,
\begin{equation}
	\lambda_B=k g_{L} \lambda_0^2 B,
\end{equation}
where the symbols are those defined for Eq.~(\ref{g_eff}).
Since the magnetized plasma may 
not completely fill each resolution
element, the Stokes $I$ and $V$ profiles must include 
a fraction $(1-\alpha)$ produced by the unmagnetized plasma in the resolution
element,
\begin{eqnarray}
V(\lambda)=&\alpha V_m(\lambda),\cr
I(\lambda)=&\alpha I_m + (1-\alpha) I_q(\lambda).
	\label{quiet}
\end{eqnarray}
The solid line in Fig.~\ref{calibration} has been computed using these
equations, together with the quiet Sun profiles
$I_q$ observed at 
disk center \citep[Brault \& Neckel 1987, quoted in][]{nec99}.
We vary the longitudinal magnetic
field strength $B$ along the sequence, assuming 
$\alpha B$ to be constant (30 G; a 
value similar to the effective magnetic field
in our magnetograms).
As expected, the ratio is one for weak fields and 
increases to 1.4 for $B$ of the order of 1.5 kG.
Figure \ref{example} illustrates
the reason of such behavior.
It shows  the two terms of the magnetograph
Eq. (\ref{eeqmag})
for weak fields ($B=150$ G)  and strong fields
($B=1500$ G). When the field is weak then
the magnetograph equation is a good approximation and
the two terms are identical. The 
two spectral lines give the true flux density and
therefore the ratio becomes one.
On the other and,
the magnetograph equation is no longer  
a good representation of the Stokes $V$ when
the field is strong. The Stokes $V$ profiles
are  much too broad as compared with the derivative of the Stokes
$I$ profiles. (Note that Stokes $I$ is 
almost exclusively produced
by unmagnetized plasma, since $\alpha\ll 1$ 
in  Eq.~(\ref{quiet}).) This so-called {\em saturation}  biases the 
effective magnetic field below the real flux density. Since the
effects increases with increasing  $g_{L}$,
$B_{eff}(6301)/B_{eff}(6302) \ge 1.$

	In order to show that the above calibration is independent
of the thermodynamic parameters of the atmosphere, we repeat the
calculations using synthetic profiles from a hot model
atmosphere (Solanki 1986 network, with 
constant magnetic field, constant microturbulence of 1 km\,s$^{-1}$,
and a macroturbulence of 1 km\,s$^{-1}$).
The dependence
of the ratio on the field strength is also represented in Fig.
\ref{example} (the long dashed line). The differences
with respect to the quiet Sun atmosphere are 
small and not discussed further.

	We use a third  type of model atmosphere to
calibrate the ratio, namely, the model MISMAs 
by \citet{san00}. These semi-empirical atmospheres are able
to account for all Stokes profiles of \lineone\ and
\linetwo\ observed in the quiet Sun at the disk center,
including their asymmetries.
The magnetic field strength of the
model MISMAs
varies with height in the atmosphere. We modify the
field strength of the original models by changing the
magnetic field strength at the base of the 
atmosphere, and then recomputing the full vertical
stratification.
The mean magnetic field of the atmosphere
represented in Fig. \ref{calibration} is
chosen as 
the magnetic field strength at the height where
the line core opacity of \linetwo\ equals one.
The variation of the synthetic ratio versus mean
magnetic field strength is qualitatively
different than the variation found in the 
two previous cases, however, the difference is 
significant  only
when the magnetic field is weak.
The observed ratio (\ref{ratio0})
also requires  kG magnetic field strengths.
The reason why the ratio 
$B_{eff}(6301)/B_{eff}(6302)$
does not tend to one 
for weak field has to do with the stray-light
contamination of the 
model MISMAs (a contribution
equivalent to the term $[1-\alpha] I_q$ in Eq.~(\ref{quiet})).
It has a 
velocity distribution very different from the velocities
of the magnetized plasma. Then the
magnetized and un-magnetized atmospheres are no longer
similar, which causes the break down of the magnetograph
Eq. (\ref{eeqmag}). Figure \ref{calibration} includes
several kinds of MISMAs, from simple common ones
(classes 0 and 2) to 
atmospheres having opposite polarities
in the resolution element (classes 6 and 7).
All of them assign similar magnetic field strengths to
the observed ratio (\ref{ratio0}).

\end{document}